\documentclass[twocolumn]{aastex61}

\shorttitle{Giant planets' configuration and evolution}
\shortauthors{Deienno et al.}

\begin{document}


\title{Constraining the giant planets' initial configuration from their evolution: \\ implications for the timing of  the planetary instability}

\author{Rogerio Deienno}
\affil{Laboratoire Lagrange, UMR7293, Universit\'e C\^ote d'Azur, CNRS, Observatoire de la C\^ote d'Azur, Boulevard de l'Observatoire, 06304 Nice Cedex 4, France}
\affil{Instituto Nacional de Pesquisas Espaciais, Avenida dos Astronautas 1758, CEP 12227-010 S\~ao Jos\'e dos Campos, SP, Brazil}
\correspondingauthor{Rogerio Deienno}
\email{rogerio.deienno@gmail.com}

\author{Alessandro Morbidelli}
\affil{Laboratoire Lagrange, UMR7293, Universit\'e C\^ote d'Azur, CNRS, Observatoire de la C\^ote d'Azur, Boulevard de l'Observatoire, 06304 Nice Cedex 4, France}

\author{Rodney S. Gomes}
\affil{Observat\'orio Nacional, Rua General Jos\'e Cristino 77, CEP 20921-400 Rio de Janeiro, RJ, Brazil}

\author{David Nesvorn\'y}
\affil{Department of Space Studies, Southwest Research Institute, 1050 Walnut St., Boulder, CO 80302, USA}

\begin{abstract}
Recent works on planetary migration show that the orbital structure of the Kuiper belt can be very well reproduced if before the onset of the planetary instability Neptune underwent a long-range planetesimal-driven migration up to $\sim$28 au.
However, considering that all giant planets should have been captured in mean motion resonances among themselves during the gas-disk phase, it is not clear whether such a very specific evolution for Neptune is possible, nor whether the instability could have happened at late times.
Here, we first investigate which initial resonant configuration of the giant planets can be compatible with Neptune being extracted from the resonant chain and migrating to $\sim$28 au before that the planetary instability happened. 
We address the late instability issue by investigating the conditions where the planets can stay in resonance for about 400 My. 
Our results indicate that this can happen only in the case where the planetesimal disk is beyond a specific minimum distance $\delta_{stab}$ from Neptune. 
Then, if there is a sufficient amount of dust produced in the planetesimal disk, that drifts inwards, Neptune can enter in a slow dust-driven migration phase for hundreds of Mys until it reaches a critical distance $\delta_{mig}$ from the disk. 
From that point, faster planetesimal-driven migration takes over and Neptune continues migrating outward until the instability happens. 
We conclude that, although an early instability reproduces more easily the evolution of Neptune required to explain the structure of the Kuiper belt, such evolution is also compatible with a late instability.
\end{abstract}


\section{Introduction}
It is now well accepted that the giant planets of the Solar System did not have the current orbits when the gas was removed from the protoplanetary disk.
The first work showing that the planets had to migrate while interacting with the remaining planetesimal disk was \citet{fernandez1984}. 
This work demonstrated that, in the planetesimal scattering process, on average Saturn, Uranus, and Neptune gain angular momentum and thus migrate outwards, while Jupiter loses angular momentum and moves inwards. 
Since then, the process of planetesimal-driven migration was considered to be a fundamental one in shaping the current structure of the solar system. 
\cite{malhotra1993,malhotra1995} showed that planetesimal-driven migration can explain the origin of the orbit of Pluto and of the resonant populations of Kuiper belt objects. 
However, it appeared soon that planetesimal-driven migration alone is a process too smooth to explain several aspects of the Kuiper belt's orbital distribution, e.g. the orbital excitation of the non-resonant objects. 
\citet{thommes1999} was the first paper showing that a dynamical instability of the giant planets could lead to a global orbital excitation of the Kuiper belt, while delivering the giant planets on final orbits roughly similar to the current ones. 

Building on these results, the so-called {\it Nice model} \citep{tsiganis2005,morbidelli2005,gomes2005} combined the processes of planetesimal-driven migration and dynamical instability into a new global scenario. 
It suggested that the planets Jupiter, Saturn, Uranus, and Neptune formed in a compact configuration with Jupiter around 5.45 au, Saturn slightly interior to the 2:1 mean motion resonance (MMR) with Jupiter ($a_{2:1}$ $<$ 8.65 au), and the initial semimajor axes of the ice giants (Uranus and Neptune) in the ranges 11--13 au and 13.5--17 au. 
The planets were surrounded by a disk of leftover planetesimals, accounting for a total of about 35 $M_{\oplus}$, from $\sim$16 to 30 au (where $M_{\oplus} \sim 5.97 \times 10^{24} ~\rm kg$ is the mass of the Earth). 
In this model the planets interacted with the planetesimal disk and, by planetesimal-driven migration, their mutual separations slowly increased. When Jupiter and Saturn crossed their mutual 2:1 MMR  the full system of the giant planets became unstable and spread further through mutual close encounters. 
In particular, Uranus and Neptune were scattered onto orbits with relatively large eccentricities and semimajor axes, which crossed the original trans-Neptunian disk. 
Then, following a phase of dynamical friction caused by their interaction with the disk, the orbits of Uranus and Neptune damped in eccentricity and inclination. 
A residual planetesimal-driven migration placed the planets on orbits fairly similar to the current ones \citep{tsiganis2005}.

The Nice model had great success in explaining many features of our Solar System, as: the capture of Jupiter's Trojans \citep{morbidelli2005}, the capture of the irregular satellites of the giant planets \citep{nesvorny2007}, the absence of regular moons beyond Oberon's orbit at Uranus \citep{deienno2011}, to cite a few. 
Moreover, it was shown that, provided the inner edge of the trans-Neptunian disk was at an appropriate distance from the original orbit of Neptune, the giant planet instability could have occurred after hundreds of My of evolution, potentially explaining the origin of the so-called Late Heavy Bombardment (LHB) of the terrestrial planets \citep{gomes2005}. 

However, the original version of the Nice model  presented a major problem; the initial configuration of the planets was chosen in an ad hoc manner.  
\citet{morbidelli2007} performed hydrodynamical simulations of the evolution of the giant planets when they were still embedded in a disk of gas and found that the planets should have acquired a fully resonant configuration, with Jupiter and Saturn preferentially locked in their mutual 3:2 MMR with Jupiter at $\sim$5.4 au, Saturn and Uranus (ice I) in the 3:2 MMR, and Uranus (ice I) and Neptune (ice II) in the 4:3 MMR (although other resonant configurations are also possible). 
The authors also concluded that such a multi-resonant system is compatible with a subsequent evolution similar to that described in \citet{tsiganis2005}, although the instability in this case happens when a pair of planets leave the original resonant configuration, rather than when Jupiter and Saturn cross their 2:1 MMR. 

\citet{levison2011} investigated whether, under some conditions, this initial multi-resonant configuration could still lead to a {\it late} giant planet instability, in order to explain the putative LHB as in the original \citet{gomes2005} paper. 
They found that late instabilities are possible, provided that the inner edge of the disk is several au's beyond the outermost planet and about 1,000 Pluto-sized planetesimals are embedded in the disk, so to produce an important dynamical self-stirring of the planetesimals population.

During the dynamical instability of the giant planets, several dynamical paths are possible, even when restricting the evolutions to those leading to final planetary orbits similar to the current ones. 
The actual dynamical evolution of the giant planets at the instability time can be constrained by investigating the consequences on the other components of the Solar System: the terrestrial planets and the small body populations. 

The terrestrial planets and the asteroid belt give similar constraints, as investigated in \citet{brasser2009} and \citet{morbidelli2010}. 
If the orbits of Jupiter and Saturn migrate away from each other smoothly and slowly, as in planetesimal-driven migration, the $g_5$ secular frequency (dominant in the precession of Jupiter's longitude of perihelion) decreases slowly in value and becomes very similar for some time to the $g_1$ secular frequency, dominant in the precession of Mercury's perihelion. 
As a consequence, the orbit of Mercury is destabilized \citep{brasser2009}. 
Similarly, the $g_6$ secular frequency, dominant in the precession of Saturn's perihelion, as it slows down becomes equal to the frequencies characterizing the precession of the asteroids at low inclination in the inner asteroid belt, so that this region becomes severely depleted  \citep{morbidelli2010}. 
Both papers concluded that the ratio of the orbital periods of Jupiter and Saturn $(P_s/P_j)$ had to evolve almost discontinuously from $< 2.1$ to $> 2.3$ (the current value is  $P_s/P_j \approx 2.49$). 
This happens if Jupiter and Saturn encounter another planet (Uranus or Neptune or a third planet of comparable mass). 
The sequence of encounters causes a divergent jump in the semimajor axes of the orbits of Jupiter and Saturn, and therefore a jump in the $P_s/P_j$ ratio. 
This scenario is commonly known as the {\it jumping-Jupiter} model. 
Notice that if the giant planet instability occurred early, the terrestrial planets were possibly not yet fully formed and therefore the constraint on the evolution of $g_5$ relative to $g_1$ does not apply \citep{kaib2016}. 
However, the constraint set by the asteroid belt is still valid, supporting the need for a {\it jumping-Jupiter} evolution regardless of the time of the instability \citep{walshmorby2011,toliou2016}.

Given that in a {\it jumping-Jupiter} evolution the planet that encounters Jupiter is often ejected onto a hyperbolic orbit, \citet{nesvorny2011} and \citet{batygin2012} suggested that the Solar System had originally a third ice giant planet with a mass comparable to Uranus or Neptune. 
This scenario increases the probability that a {\it jumping-Jupiter} evolution ends with four planets near their current orbits, once the putative fifth planet is ejected after its encounter with Jupiter. 

Then, \citet{nesvorny2012}  considered many different initial multi-resonant configurations for Jupiter, Saturn, Uranus, Neptune, and the rogue fifth planet, and determined for each of them the probability that the $P_s/P_j$ ratio had a jump across the 2.1--2.3 range and the final planets reached orbits compatible with those of our current Solar System. 
The successful simulations of \citet{nesvorny2012} have then been used to investigate again the capture of the Trojans of Jupiter \citep{nesvorny2013}, the capture of irregular satellites at Jupiter \citep{nesvorny2014}, the excitation of the inclination of the moon Iapetus at Saturn \citep{nesvorny2014b}, the evolution of the asteroid main belt and of its collisional families \citep{roig2015,deienno2016,brasil2016}. 
They have also been shown to be compatible with the survival of the Galileans satellites \citep{deienno2014}, despite of the encounters of Jupiter with the ejected planet. 

However, when investigating the dynamical sculpting of the Kuiper belt,  \citet{nesvorny2015a,nesvorny2015b} and \citet{nesvorny2016} realized that, of all possible evolutions of Neptune consistent with the five-planet {\it jumping-Jupiter} model, only some are consistent with the current orbital structure of the trans-Neptunian population. 
In order to get the correct inclinations of the hot population of the Kuiper belt objects (KBO), \citet{nesvorny2015a} found that Neptune should have migrated more that 5 au from its original resonance by planetesimal-driven migration before that the planets became unstable. 
This migration should have occurred on a timescale $\tau$ $\geq$ 10 My, with the planet on a quasi-circular planar orbit ($e_{_N} < $ 0.1 and $i_{_N} < 2^\circ$) to avoid excessive orbital excitation of the cold classical belt, in the $\sim$42--45 au region. 
The planetary instability should have happened when Neptune was already at 28 au \citep{nesvorny2015b} in order to explain the so-called {\it Kernel} of the cold Kuiper-belt (a clump of objects around 44 au; \citep{petit2011}). 
In this case the {\it Kernel} would have formed from objects transported outwards in the 2:1 MMR with Neptune and released from the resonance when Neptune's orbit had a jump of about 0.5 au in semimajor axis, due to a close encounter with another planet. 
Finally, in order to reproduce the observed ratio between the resonant and non-resonant Kuiper belt populations, \citet{nesvorny2016} concluded that the planetesimal-driven migration of Neptune should have been characterized by several small amplitude jumps, as those caused by a population of 1,000--4,000 Pluto-size objects in the disk. 

The brief review reported above, sets the context that motivated the study reported in this paper. 

It is clear that the work by \citet{nesvorny2012} needs to be revisited to test which initial resonant configurations are the most compatible with the new constraints on the dynamical evolution of Neptune. 
This is done in the first part of the paper (section~\ref{early}). 
In the second part of the paper (section~\ref{late}) we investigate whether the initial configurations selected in the first part of the work can lead to a {\it late} planetary instability, while still showing an evolution of Neptune consistent with Kuiper belt constraints. 

The possibility of a {\it late} instability is far from trivial. 
In fact, in order to exhibit a long-range planetesimal-driven migration before the planet instability, Neptune needs to be embedded in the planetesimal disk. 
But in this case, Neptune would start migrating away from the other planets from the very beginning of the simulation (corresponding to the gas-removal time). 
This is likely to lead to an instability after just a few 10s of My. 
If instead the planetesimal disk is relatively far from Neptune, the planet can be extracted from its initial resonance at a {\it late} time as in \citet{levison2011}, but the instability is likely to happen as soon as the initial resonant configuration is broken, i.e. before that Neptune can migrate far away. 
For this problem we will introduce a new concept: that of dust-driven migration, i.e., migration due to the gravitational interaction of a planet with the dust produced in a distant planetesimal disk, spiraling inward by Poyning-Robertson drag. 
We will study the evolution of the system as a function of the dust properties and the numerical resolution used in modeling the planetesimal disk. 
Finally, our results are summarized and discussed in section \ref{conclusions}.

\section{Initial configuration of the giant planets and the constraints in the Kuiper Belt}\label{early}

\citet{nesvorny2012} characterized the evolution of the giant planets from several possible initial resonant configurations. 
They assumed that the Solar System could have formed with 4, 5, or 6 giant planets. 
The additional planets had a total mass comparable to that of Uranus or Neptune. 
They analyzed the rate of success of each configuration by measuring the fraction of the planetary orbital evolutions that fulfilled a set of constraints, namely: the final planetary system is made of four giant planets (criterion A), the orbits of the 4 remnant planets resemble the present ones (criterion B), the proper mode of Jupiter's eccentricity ($e_{_{55}}$) is greater than 0.022 in the final system, i.e., at least half of its current value (criterion C), and the ratio $P_s/P_j$ evolves from $< 2.1$ to $> 2.3$ in less than 1 My (criterion D). 
Criterion C is important because it is found that, although encounters between Jupiter and an ice giant can excite $e_{_{55}}$, not all of them can do it as much as needed. Criterion D is important because it is linked with the evolution of the terrestrial planets and of the asteroid belt \citep{brasser2009,morbidelli2010,toliou2016}. If Jupiter and Saturn migrate too slowly through the period ratio 2.1--2.3, the $g=g_5$ and $g=g_6$ resonances sweep through the terrestrial planet region and the inner asteroid belt, resulting in orbital distributions inconsistent with those observed. Then, from the success rates, \citet{nesvorny2012} concluded that a planetary system initially having 5 planets should be preferred. 

Here, we revisit the \citet{nesvorny2012} work, adding a new constraint on the orbital evolution of Neptune from the work of  \citet{nesvorny2015b}, namely that Neptune should have migrated to $\sim$28 au before the onset of the planetary instability. 
The works by \citet{nesvorny2015a,nesvorny2015b} and \citet{nesvorny2016} also give other constraints on the migration timescale of Neptune and its eccentricity and inclination before the instability. 
The eccentricity and inclination constraints are met fairly easily because in this scenario Neptune evolves in a planetesimal-driven migration without planetary close encounters until the instability phase. 
As for the migration timescale, \citet{nesvorny2015a} claims that before the planetary instability, once out of resonance, Neptune can migrate according to $\tau =$ 10 My for disks with 20 $M_{\oplus}$ and according to $\tau =$ 50 My for disks with 15 $M_{\oplus}$. Considering that those disks have the same extension, this exemplifies how sensitive the migration timescale can be with respect to the density of the disk.
Therefore, we decide not to consider the migration timescale as an important parameter at this point, because getting a correct timescale of the migration is probably only a matter of tuning the density and profile of the planetesimal disk. 
Thus we add just one new criterion (criterion E): at the time of the instability $a_{_N}$ has to be in the interval 27--29 au. 
While criteria A to D indicate that the initial Solar System most probably hosted five giant planets \citep{nesvorny2012}, criterion E is a very important criterion that can be used to constrain in which resonant chain these five planets likely were at the end of the gas-disk phase.

For simplicity, we consider only the cases of \citet{nesvorny2012} that involved 5 planets.  
We also always considered Jupiter initially $\sim$5.4 au, a distance between the outermost ice giant and the inner edge of the planetesimal disk $\delta  =$ 1 and 2 au 
and extending to 30 au \citep{gomes2004}, composed by 1,000 planetesimals (which is the resolution considered by \citet{nesvorny2012}), and two values of surface density ($\sum(r)=1/r$) equal to 0.02 and 0.03 $M_{\oplus}/au^2 $. 
Such configuration implies that the mass of the planetesimal disk ranges from $\sim$13 $M_{\oplus}$ (in the more relaxed configuration -- 3:2, 3:2, 2:1, 2:1\footnote{In this paper we denote a multi-resonant configuration by a sequence of integer ratios, 
identifying the MMRs between adjacent planets, Jupiter and Saturn
always being the first and the second from the Sun with $a_{_J}\sim$ 5:4 au, and the three equal-mass ice-giant planets placed beyond them}, with $\delta =$ 2 au and $\sum(r)=$ 0.02 $M_{\oplus}/au^2 $) to $\sim$65 $M_{\oplus}$ (in the more compact configuration -- 3:2, 3:2, 4:3, 5:4, with $\delta =$ 1 au and $\sum(r)=$ 0.03 $M_{\oplus}/au^2 $). 

To perform our simulations, we used the Hybrid integrator in the {\it Mercury-package} \citep{chambers1999}, with a time step of half an year. 

Table \ref{table1} gives in column 6 the rate of success for our new criterion E, after 20 runs of each initial configuration from \citet[][table 6 in their paper]{nesvorny2012}. 
Those 20 simulations were composed by 5 simulations for each pair $\delta=$ [1,2] au, and $\sum(r)=$ [0.02-0.03] $M_{\oplus}/au^2 $.
Although we have used two different values of surface density and two values of $\delta$ for the planetesimal disk, we decided to compress the results of all 20 simulations, for each configuration, in table \ref{table1}, column 6. 
This is because, concerning our criterion E, only minor differences were observed with these changes. 
Thus, when important, those minor differences will be reported.
For completeness and to increase statistics, table \ref{table1} also reports the average of the rate of success for the other criteria taken from \citet[][table 6 in their paper]{nesvorny2012}.
Because the simulations performed in this work are essentially the same of \citet{nesvorny2012}, the success rate for criteria A to D are consistent (within the small number statistics) with those reported in \citet{nesvorny2012}. Thus, we did not report our own results for the criteria A, B, C, and D.
Rather than that, we decided to check only the success rate relative to the new criterion E, and then, combine our findings with those of \citet{nesvorny2012}, which have much higher statistics, in order to determine which of the initial planetary configurations are the most compatible with the constraint on the dynamical evolution of Neptune, unveiled in \citet{nesvorny2015a,nesvorny2015b}.

\begin{deluxetable}{lccccc}
\tablecaption{Success rates of five giant-planet simulations. Columns (1) to (5) are from \citet{nesvorny2012}: (1) Initial configuration of the planets. (2)--(5) average of the success rates for the four criteria taken from \citet{nesvorny2012}. Column (6) reports the success rate for our new criterion, i.e., 27~au~$\le$ $a_{_N}$ $\le$~29~au at the time of the instability. In boldface we highlight the initial configuration that has the highest success rate for all the constraints together. \label{table1}}
\tabletypesize{\scriptsize}
\tablehead{
\colhead{Initial Configuration} & \colhead{A (\%)} & \colhead{B (\%)} & \colhead{C (\%) } & \colhead{D (\%)} & \colhead{E (\%)}
}
\startdata
3:2, 3:2, 4:3, 5:4     &     24    &     12     &       0     &      2     &    5  \\
3:2, 3:2, 3:2, 3:2     &     30    &     15     &       2     &      5     &    15  \\
3:2, 3:2, 4:3, 4:3     &     47    &     23     &       3     &      3     &     5  \\
{\bf 3:2, 3:2, 2:1, 3:2} & {\bf 32} & {\bf 13} & {\bf  4} & {\bf 7} & {\bf  65} \\
3:2, 3:2, 2:1, 2:1     &     38    &     18     &       4     &      4     &    25   \\
2:1, 3:2, 3:2, 3:2     &     53    &     40     &       2     &     18    &     0    \\
2:1, 4:3, 3:2, 3:2     &     73    &     33     &       0     &     18    &    40  \\
\enddata
\end{deluxetable}

Our results can be interpreted as follows:

\underline{{\it 2:1Jupiter--Saturn resonance}}: in these configurations, as it also appears in \citet[][Fig. 16]{nesvorny2012}, it is possible that Neptune leaves the resonant configuration and migrates ahead the other planets, but not always in the way required in \citet{nesvorny2015a,nesvorny2015b}. 
Indeed, a configuration with initial resonances 2:1, 3:2, 3:2, 3:2 never lead to an instability that fulfilled criterion E in our 20 simulations. 
The evolutions from the 2:1, 4:3, 3:2, 3:2 configuration, on the other hand, had one of the largest rate of success for criterion E. 
However, such configurations present other major problem, mostly concerning the final orbital separation between Jupiter and Saturn, which is typically too big. 
The point is that, because the ratio $P_s/P_j$ has to change less than 0.5 if Jupiter and Saturn are initially in the 2:1 MMR, the mass of the fifth planet should be $\sim$1/2 of the mass of Neptune. Instead, we assumed for this planet a mass equal to that of Uranus-Neptune, as in \cite{nesvorny2012}. 
A smaller mass planet, however, would not excite Jupiter's eccentricity enough, so that the criterion C would not be fulfilled. 
Moreover, hydrodynamical simulations \citep{morbidelli2007,pierens2008} show that in a massive gas-disk, Jupiter and Saturn would be locked in their mutual  3:2 MMR, rather than 2:1. The Grand Tack model \citep{walsh2011} also implies that Jupiter and Saturn were in their 3:2 MMR at the disappearance of the disk of gas. So, we don't consider that configurations with Jupiter and Saturn in the 2:1 MMR are very realistic. 

\begin{figure*}
\gridline{\fig{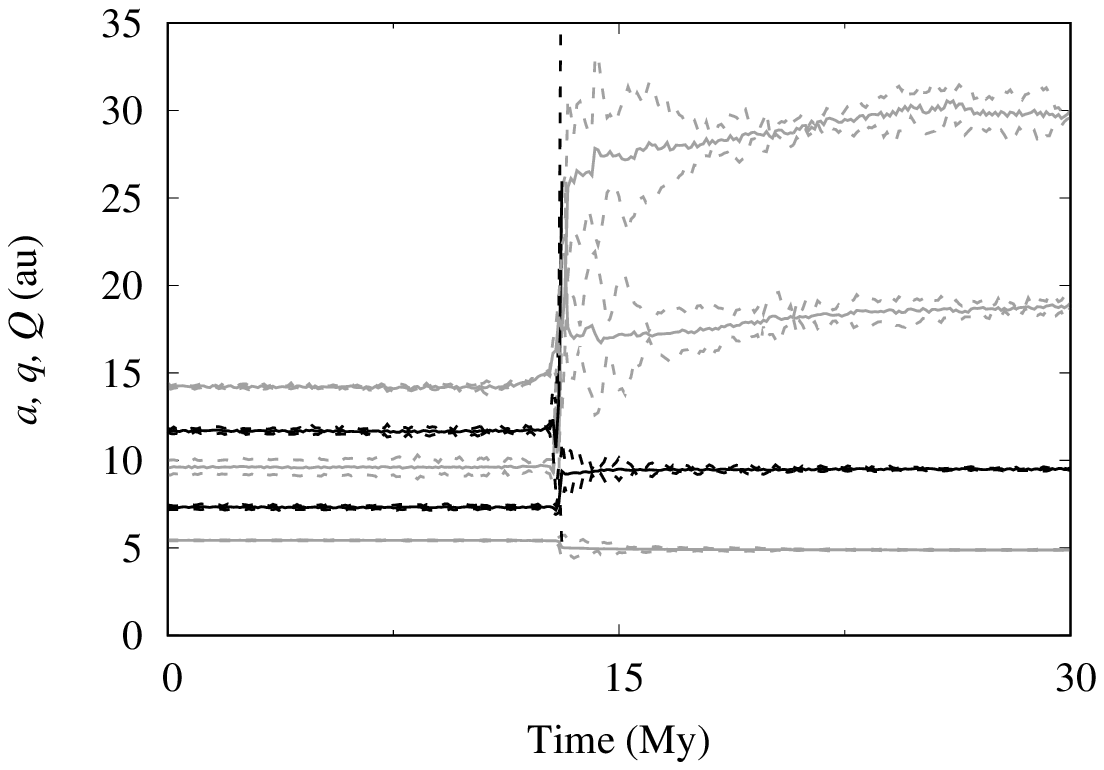}{.5\textwidth}{}
          \fig{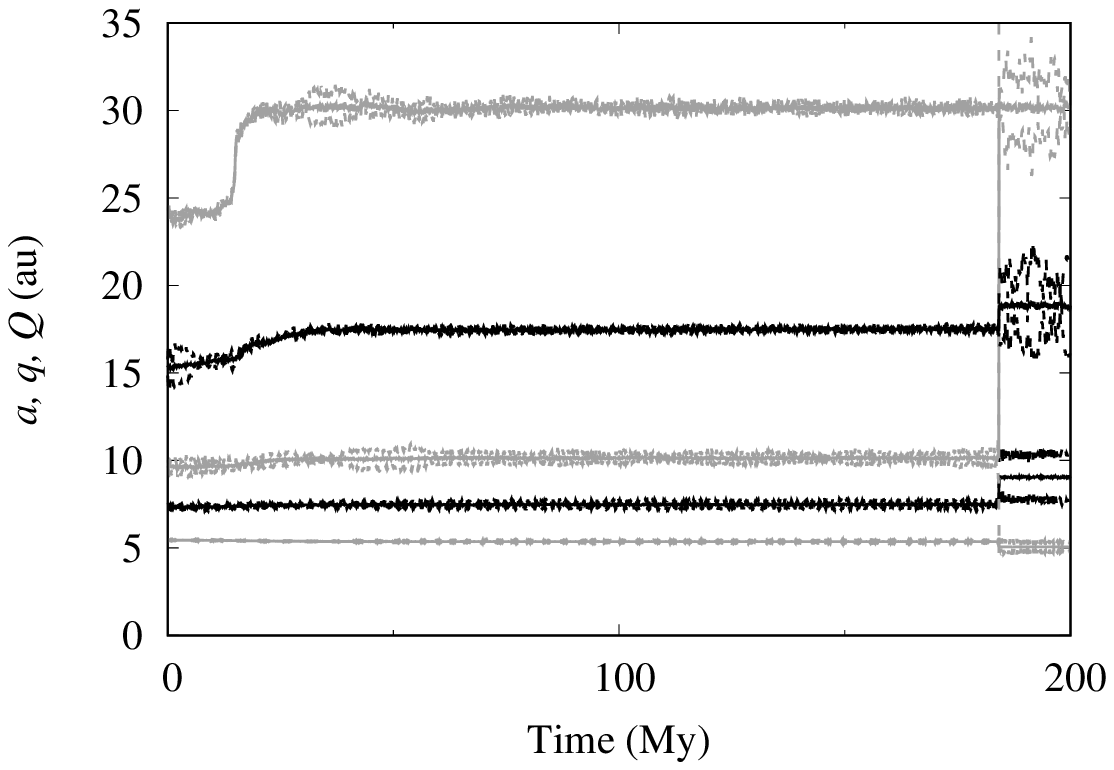}{.5\textwidth}{}}
\gridline{\fig{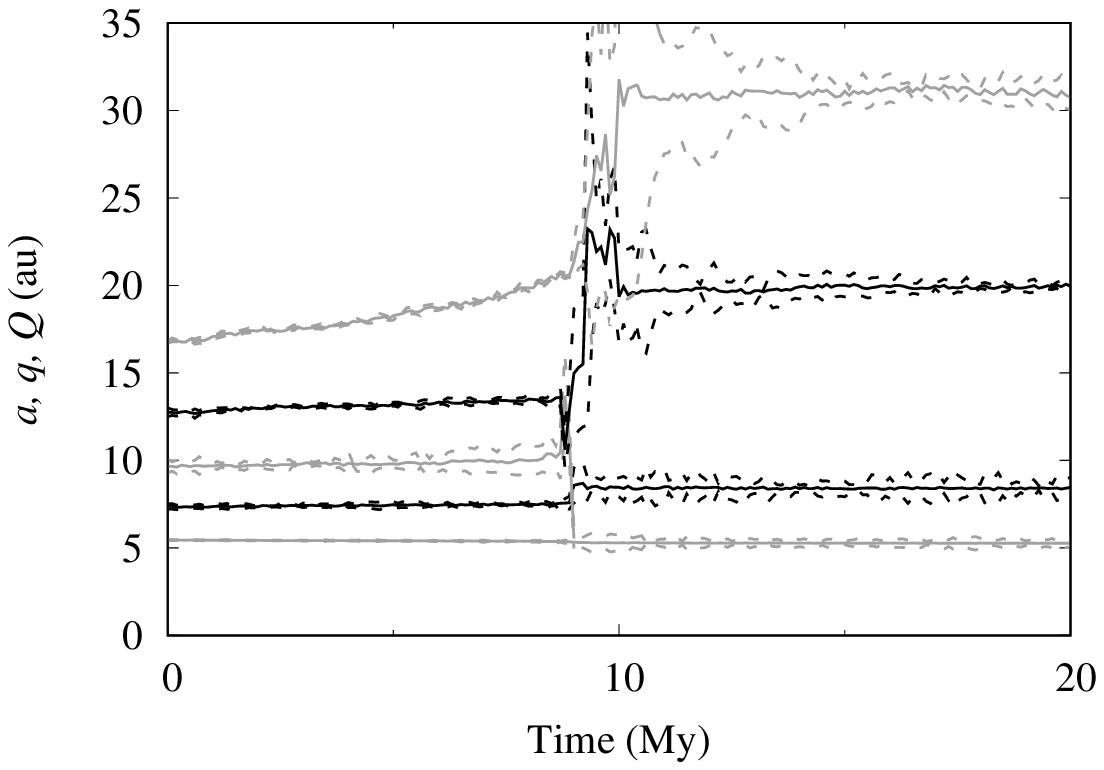}{.5\textwidth}{}
          \fig{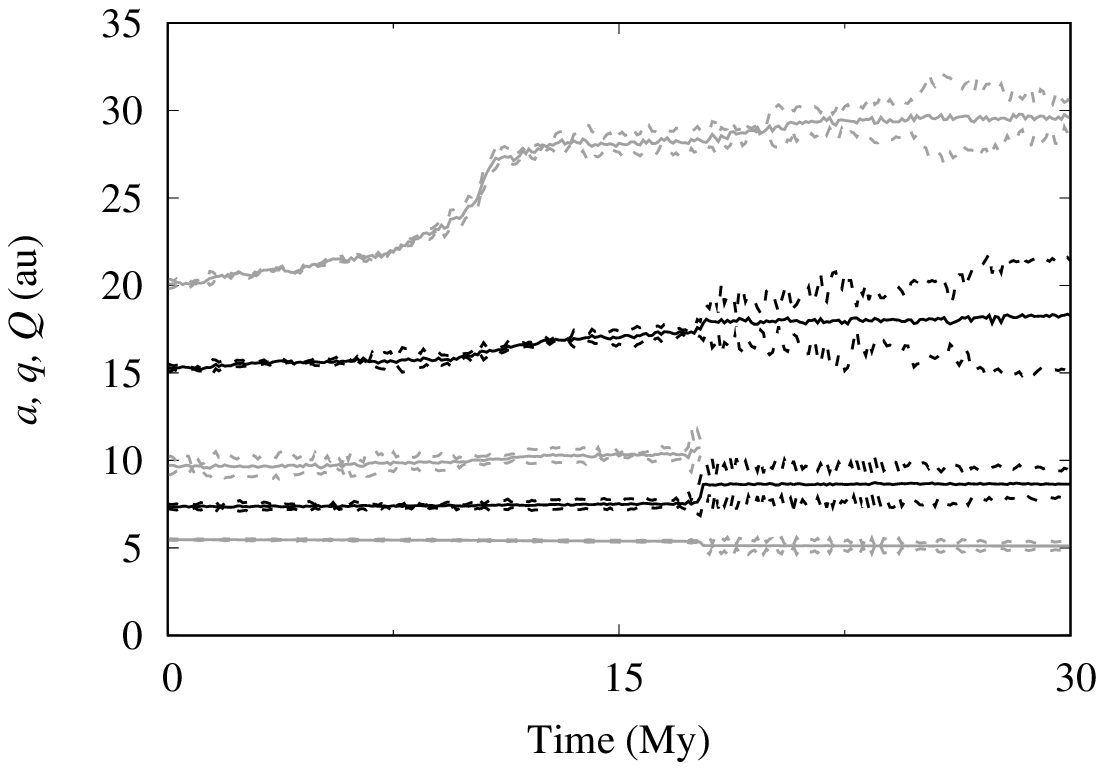}{.5\textwidth}{}}
    \caption{
Semimajor axis ($a$), perihelion distance ($q$) and aphelion distance ($Q$) for all planets as a function of the time. 
The initial configurations are:
Top left panel: 3:2, 3:2, 4:3, 4:3 (criteria fulfilled: A, B, and D). 
Top right panel: 3:2, 3:2, 2:1, 2:1 (criteria fulfilled: A, B, C and D).  
Bottom left: 3:2, 3:2, 3:2, 3:2 (criteria fulfilled: A, B, C and D).  
Bottom right panel: 3:2, 3:2, 2:1, 3:2 (criteria fulfilled: A, B, C, D and E).  
All configurations start with Jupiter at $\sim$5.4 au.
Disk's parameters and criteria are explained in the main text.
}
    \label{fig1}
\end{figure*}

\underline{{\it 3:2, 3:2, 4:3, 4:3(5:4)}}: these two configurations are too compact. 
Only one out of 20 simulations for each configuration fulfilled criterion E.
Usually, the evolution of the planetary system is as in \citet[][Fig. 7]{nesvorny2012}, where the instability takes place as soon as one planet goes out of resonance. As a result, no large-range planetesimal-driven migration occurs before the planetary instability. Instead, the instability is violent and the evolution of Neptune is largely different from that constrained in \citet{nesvorny2015a,nesvorny2015b} (figure \ref{fig1} top left, in this work). 
Nevertheless, although they are low, none of the success rates reported in table \ref{table1} for the configuration 3:2, 3:2, 4:3, 4:3 are null. Thus, even being very unlikely, due to the reasons presented, this configuration cannot be ruled out.

\underline{{\it 3:2, 3:2, 2:1, 2:1}}: this configuration is too loose. 
\citet{nesvorny2012} already pointed out that Neptune stops farther than 30 au for whatever reasonable disk's parameters starting for this configuration. 
Here, we found that even with the planetesimal disk of lowest mass ($m_{disk} \sim$ [13--24] $M_{\oplus}$), ranging from $a_{disk} \sim$ [25(26)--30] au, Neptune goes out of resonance and usually migrates all the way up to 30 au or more, before that the planetary instability happens (figure \ref{fig1} top right). However, sometimes the instability can also occur before Neptune reaches 30 au, say at $\sim$28 au. Therefore, although this configuration may also present some problems with the extension of the planetesimal disk, as it will be discussed in the following sections, it does have a good rate of success in fitting criterion E (25\% success rate). 

\underline{{\it 3:2, 3:2, 3:2, 3:2}}: this configuration allows Neptune to migrate away from the other planets before the instability, but usually not enough. 
The instability commonly happen when Neptune is around 20 au \citep[see][Figs 8 and 9, and figure \ref{fig1} bottom left in this work]{nesvorny2012}. 
Only for high density disks Neptune can migrate fast enough and reach $\sim$28 au before the planetary instability.
If the disk's density is low, Neptune migrates slowly and can't go too far into the planetesimal disk. 
So, this configuration has a low, but non-negligible, success rate of 15\% and cannot be ruled out.

\underline{{\it 3:2, 3:2, 2:1, 3:2}}: this is the configuration that presents the highest rate of success, considering all criteria together, so that it is highlighted in bold in table \ref{table1}.  
This is also the most successful configuration relative to criterion E, with a success rate of 65\%  independently of the disk setup.
The evolution of Neptune and of the rest of the planetary system is usually similar to that shown in \citet[][Figs 12, 13, and 14]{nesvorny2012}, and satisfies the constraint identified in \citet{nesvorny2015a,nesvorny2015b} (figure \ref{fig1} bottom right in this work). 

Figure \ref{fig1} shows examples of the evolution of the systems corresponding to the descriptions given for each of the resonant chains discussed above (with the exception of the 2:1 Jupiter-Saturn resonance, due to the considerations previously made). 
Almost all evolutions depicted in figure \ref{fig1} fulfill the criteria A, B, C, and D. 
Criterion C was not fulfilled only for the configuration 3:2, 3:2, 4:3, 4:3 (top left), where, although the eccentricity of Jupiter was excited during the instability, it was quickly damped afterwards ($e_{55}\sim$ 0.0075) due to the strong dynamical friction provided by the planetesimal disk.
Only the configuration 3:2, 3:2, 2:1, 3:2 (bottom right) fulfilled all constraint together, including our new criterion E.

Summarizing, when the new constraints on the evolution of Neptune from \citet{nesvorny2015a,nesvorny2015b,nesvorny2016} are considered, the most probable initial configuration of the giant planets was 3:2, 3:2, 2:1, 3:2. 
The configurations 3:2, 3:2, 2:1, 2:1, and 3:2, 3:2, 3:2, 3:2, or 3:2, 3:2, 4:3, 4;3 also appear possible, but less probable. 
This seems to be a consequence of the more relaxed orbital separation between the first and second ice giants. 
Therefore, one would expect that any configuration having the first and the second ice giant in distance from the Sun in a 2:1 MMR or similarly relaxed could perhaps also work. However, for simplicity, we decide to restrict ourselves only to the cases of \citet{nesvorny2012}, which we know the rate of success in filling criteria A, B, C, and D.

The analysis performed in this section does not take into account the timing of the planetary instability. 
With the location of the inner edge of the disk that we adopted, all evolutions are relatively fast, and lead to instabilities after a few 10s My. 
In the next section we analyze whether it is possible to get a late instability out of the 3:2, 3:2, 2:1, 3:2 configuration, under some assumption on the geometry of the disk, still satisfying criterion E on the evolution of Neptune. 

\section{Late instability}\label{late}

Since \citet{tsiganis2005} presented the {\it Nice model}, the existence of a late heavy bombardment of the terrestrial planets has been associated to the possibility that the giant planet instability occurred after hundreds of My of evolution \citep{gomes2005,bottke2012}. 
Actually, the Nice model, both in its initial version and from generic initial resonant configurations \citep{levison2011}, is compatible with both early and late instabilities, depending on the distance of the inner edge of the original trans-neptunian disk and other disk properties (mass, excitation, etc.). Thus, the Nice model does not prove the existence of a late instability. The timing of the instability can only be determined by looking at other constraints, such as the lunar impact chronology. Unfortunately, the latter is still under debate. 

However, it is not obvious that the Nice model can still be compatible with a late instability, if we fix a specific initial resonant configuration and impose a specific evolution of Neptune. As explained in the introduction, if the disk is sufficiently close to Neptune at the beginning, it is relatively easy that Neptune migrates several au away before that the instability happens, but in this case the full dynamical evolution is rapid, as in the previous section. If the initial location of the disk is sufficiently far from Neptune, the instability may happen late, but Neptune is unlikely to migrate significantly before that the instability happens. Moreover, in the case of the most relaxed resonant configurations (e.g. 3:2, 3:2, 2:1, 2:1, which has Neptune initially around $\sim$24.5 au), the requirement that the inner edge of the disk is several au beyond Neptune while the outer edge is at $\sim$30~au reduces the disk to a ring. This seems implausible, particularly when considering the constraint that the disk should have hosted 1,000-4,000 Pluto-size bodies \citep{nesvorny2016}, which would probably be problematic for the late instability as well, because the spreading of the narrow disk would not permit a delay.

If one could show that the most developed and constrained version of the Nice model is incompatible with a late instability, this would be very important to clarify the chronology of events in the Solar System. Thus, in this section we try hard to find whether the instability could have happened late and under which conditions. We let the reader judge whether these conditions are realistic or plausible, with our considerations summarized in Sect.~\ref{conclusions}.  

For this goal, we proceed in steps. First, in sect.~\ref{dr}, we determine two critical values for the distance of the inner edge of the disk from Neptune. One, $\delta_{\rm mig}$, is the maximal distance that allows Neptune to have a long-range planetesimal-driven migration away from the other planets, as in the previous section. The other one, $\delta_{\rm stab}$ is the minimal distance of the disk for which the five-planet system remains stable in resonance for all times. Clearly   $\delta_{\rm stab} > \delta_{\rm mig}$ and $\delta_{\rm mig}$ is larger than or equal to the Neptune-disk distances considered in the previous section. The problem is that both $\delta_{\rm stab}$ and $\delta_{\rm mig}$ depend on disk resolution (number of planetesimals and individual masses that are used to model the disk) and that realistic values cannot be used in practical simulations. So, we need to estimate trends to evaluate the values of the critical distances. 

In a second step, in sect~\ref{de}, we study how a flux of dust from the planetesimal disk, due to collisional planetesimal grinding and Poynting-Robertson drag, influences planet migration. This depends on total mass flux and individual dust size. The idea is that, under an appropriate dust-flux, Neptune may migrate away from its initial location, so that its distance from the planetesimal disk decreases from $>\delta_{\rm stab}$ to $<\delta_{\rm mig}$. At this point planetesimal-driven migration starts, as in the previous section. If the dust-driven migration is slow enough, the planet instability can occur late, even though it occurs only a few millions of years after the beginning of planetesimal-driven migration. 

This concept is then proven in section~\ref{dde}, where we combine in a single simulation dust-driven migration and planetesimal-driven migration,  and we obtain an instability after 300~My, with Neptune's evolution satisfying criterion E. This simulation is of course performed with an unrealistically low disk-resolution, but is a proof of concept that, for whatever disk resolution, an appropriate dust flux can be found.  

We finally discuss what a realistic dust flux could be in sect.~\ref{dgl} 

\subsection{Disk resolution $vs$ stability}\label{dr}

The interaction of a planetary system with a planetesimal disk can be very resolution-dependent, i.e., sensitive on the number of planetesimals used to described the disk in the simulation and to their individual mass.

In some cases, as in those of section \ref{early}, this problem of sensitivity on disk-resolution is not strong because the distance between the outermost planet and the inner edge of the disk is small, so that there are from the very beginning many planetesimals on unstable orbits that encounter the planet, force its migration and, consequently, bring  many new planetesimals to planet-crossing orbit. This is the reason for which \citet{nesvorny2012} found similar results for test simulations conducted with 100, 300, 1,000, 3,000, and 10,000 planetesimals in the disk (same total disk masses), so that they decided to use only 1,000 planetesimals in their final simulations.  Of course,  even in these cases, the disk resolution has an influence on how grainy or smooth the planet migration will be, but doesn't have a major influence on the amplitude and the timescale of the migration.

On the other hand, the resolution of the planetesimal disk becomes a real issue when the inner edge of the disk is far enough from the outermost planet.
In this situation, the fraction of the disk that is unstable is small, mostly limited to the volume of the strongest mean motion resonances with the planet. The resonant planetesimals develop planet-crossing eccentricities and, with close encounters, force the planet to move a bit outwards. This moves outwards the resonances as well, destabilizing new planetesimals (those which were not in resonance at the beginning but are swept by the resonant motion). Depending on whether the mass of the new unstable planetesimals is larger or smaller than the mass of the original unstable planetesimals, the planet migration will accelerate or damp \citep{gomes2004,levison2007}. In the damped migration mode, the planet eventually stops migrating while the resonances become empty. 

If the number of planetesimals is small, the planetesimals are individually more massive to represent a disk with the same total mass. Thus, the migration of the planet is more grainy because each encounter between the planet and a planetesimal causes a larger jump in the planet's semimajor axis. Of course, there are fewer jumps, so that the planet migrates, on average, by the same amount. But the more jumpy migration of the planet causes the resonances to jump around as well, thus destabilizing more mass from the disk than in the case of a smoother planet migration with the same averaged range. Because more mass is transferred from the disk to planet-crossing orbit, the migration of the planet can change, in some cases, from the damped to the accelerated mode.

With this issue in mind, we proceed in our study looking for two critical disk distances, for a series of disk resolutions. 
First, we look for the minimum distance ($\delta_{stab}$) that the inner edge of the planetesimal disk has to have from Neptune (the outermost planet for our case) in order to maintain the planets stable for at least 400 My.
Second, we search for the maximal distance ($\delta_{mig}$) that Neptune can have from the inner edge of the planetesimal disk to evolve as in section \ref{early}.
From these results we will extrapolate the values of $\delta_{stab}$ and $\delta_{mig}$ to realistic almost infinite resolution\footnote{Ideally, we should have 1,000--4,000 Plutos representing the mass of 2--8 Earth masses, and the rest in 100 km class planetesimals. So, the ``infinite'' resolution should apply only for the disk mass carried by 100 km objects}.

Once again we consider disks with fixed surface densities (scaling as $\Sigma(r)=\Sigma_0/r$), because in this case we guarantee always the same amount of mass in the planets' MMRs, independently of the resolution of the disk (apart from stochastic sampling). 
The initial distribution of eccentricity and inclination of the disk was such that $e_{disk}=0$ and $i_{disk}\le 1^{\circ}$ (the same initial distribution for all disks considered in this work). This is the simplest configuration that represents a cold planetesimal disk emerging from the end of the dissipation of the gas nebula.
Even if the disk was initially more dynamical excited not much differences in the overall evolution are expected because the MMRs inside the disk are very efficient in exciting the disk's eccentricities to planet-crosser values. So, in a very short timescale an initially cold disk would appear like an initially excited one. Finally, we consider disks with resolution of 1,000, 5,000, 10,000, 15,000, and 20,000 planetesimals.  The outer edge of the disk is fixed at 30 au. 
The distance $\delta$ between the inner edge of the disk and the planet ranges from 1 to 7 au, with increments of 1 au.

Given the great demand of CPU-time required to simulate high-resolution disks, we only considered the most successful initial configuration in section \ref{early}, with planets in the 3:2, 3:2, 2:1 and  3:2 MMRs. 
For this configuration, Neptune is initially $\sim$20 au. For each value of  $\delta$, we considered six different values of $\Sigma_0$.

\begin{table}
	\centering
	\caption{$\delta_{stab}$ as a function of $\Sigma_0$ for the less resolved planetesimal disk (1,000 planetesimals). $\delta_{stab}$: minimum distance that the inner edge of the planetesimal disk has to have from Neptune (the outermost planet for our case) in order to maintain the planets stable for at least 400 My. $\Sigma_0$: density of the planetesimal disk, scaled as $\Sigma(r)=\Sigma_0/r$.
	The mass of the planetesimal disk, $m_{disk}$, in earth masses, for each pair $\Sigma_0$--$\delta_{stab}$ for the initial configuration 3:2, 3:2, 2:1, 3:2, with Neptune at $\sim$20 au and outer disk edge at 30 au is also shown for reference.}
	\label{table2}
	\begin{tabular}{ccccccccc} 
		\hline
$\Sigma_0$ ($M_{\oplus}/au^2 $) & \vline & 0.01 & 0.02 & 0.03 & 0.04 & 0.05 & 0.06  \\
		\hline
$\delta_{stab}$ (au)  & \vline &  4 &  5  &  5 & 6  & 6 & 7 \\
		\hline
$m_{disk}$ ($M_{\oplus}$)  & \vline & 10.2 & 17.2 & 25.8 & 28.0 & 35.2 & 32.2 \\
		\hline
	\end{tabular}
\end{table}

We observed that, with decreasing resolution or with increasing disk's surface density, the inner edge of the disk has to be farther and farther away from the planet to ensure the stability of the planetary system.  
For instance, table \ref{table2} shows $\delta_{stab}$ (with a 1 au accuracy) for different values of $\Sigma_0$ in the case of the less resolved disk (1,000 planetesimals). 
Indeed, $\delta_{stab}$ ranges from 4 au for the disk with the lowest density, reaching up to 7 au in the case of $\Sigma_0=$ 0.06 $M_{\oplus}/au^2 $. 
Unfortunately, we cannot run disks for 400~My with about 15,000--20,000 planetesimals.
So, we can only judge from a few experiments that increasing the disk resolution, $\delta_{stab}$ tends to be between 3--5 au depending on $\Sigma_0$. For example, when $\Sigma_0=$ 0.02 $M_{\oplus}/au^2 $, experiments with $\delta=$ 4 au (unstable for disks with 1,000 and 5,000 planetesimals), showed stable for a resolution of 10,000 planetesimals. The same resolution makes the system unstable for $\delta=$ 3 au, with Neptune starting a fast migration at around 40 My. 
But, with the same configuration, resolutions of 5,000 and 1,000 planetesimals started the Neptune's fast migration in about 25 My and 15 My, respectively. This trend in instability time {\it vs} disk resolution suggests that disks with resolutions of 15,000 and 20,000 planetesimals could perhaps be stable also for $\delta=$ 3 au, but they were not performed due to excessive cpu cost.
However, it is clear that a system with $\delta=$ 0 has to be unstable for whatever disk resolution, so the value of $\delta_{stab}$ in the limit of infinite resolution has to exist and be positive. 
Our experiments suggest that the limit value is about 1--2 au smaller than the values of $\delta_{stab}$ in table \ref{table2}, and never significantly smaller than 3 au.

Another consideration that we can retrieve from table \ref{table2} is that, values of $\delta_{stab}$ of 4--5 au preclude giant planet configurations with large separations among the planets like 3:2, 3:2, 2:1, 2:1 (where Neptune is initially $\sim$24.5 au). In fact in this case 
the ``disk'' extension would be less than 1.5 au, which is implausible, particularly taking into account that 1,000-4,000 Pluto mass bodies had to exist in this disk \citep{nesvorny2016}. Given that in the 3:2, 3:2, 3:2, 3:2 and 3:2, 3:2, 4:3, 4:3 configurations Neptune is unlikely to migrate far enough before that the instability happens (if the migration rate is slow, sect.~\ref{early}), the 3:2, 3:2, 2:1, 3:2 configuration is left to be as the only one with a chance to be compatible with a {\it late instability}. This is the configuration we are considering for the experiments in this section, and is also the one with the higher rate of success in filling criterion E (table \ref{table1} in section \ref{early}).

From another and similar set of simulations, where we restricted ourselves to $\Sigma_0=$ 0.02 and 0.03 $M_{\oplus}/au^2 $, we ran cases with 1,000, 5,000, 10,000, 15,000 and 20,000 planetesimals, observing Neptune's fast migration (as in section \ref{early}), for all resolutions when $\delta \leq$ 2 au.
For $\delta \ge$ 3 au, the evolution of the planet becomes resolution dependent, i.e., it suffers long-range migration if the disk is under-resolved, but not if the disk is modeled with tens of thousands of planetesimals (as previously described). 
Thus, we define $\delta_{mig} =$ 2 au as the distance that Neptune needs to be from the inner edge of the disk to behave as in section \ref{early} simulations independently of the resolution of the planetesimal disk. 

In order to achieve a late instability of the planetary system, we need now to find a mechanism that can slowly migrate Neptune from a distance $\delta$ from the disk larger than $\delta_{stab}$ to within $\delta_{mig}$. For this purpose, we investigate below the effects on planet migration of a flow of dust through Neptune's orbit, produced by the slow collisional grinding of the disk population. 

\subsection{Dust flux and its effect}\label{de}

In this section, we simulate  different fluxes of dust, with different mass flows and different velocity of inward migration. The latter is characterized by the parameter $\beta$, defined by the ratio between the radiation and gravitational forces, and directly depends on the size of the dust grains \citep{burns1979}. As usual in dust simulations, we will use super-particles, i.e., bodies that have the value of $\beta$ of a small dust grain but carry an important amount of mass $m_{dust}$, like a collection of many real dust grains. 

We started our analyses considering fixed values of $\beta =$ 0.1, 0.05, and 0.01, corresponding with dust grain radii of about 2.3, 4.5, and 17.0 $\mu m$, respectively \citep{moro2013}. 
Values of $\beta$ $>$ 0.1 will not be considered because those values are close enough to the limit of ejection $\beta = 0.5$. So, we expect their effects on the planets to be negligible.  
On the other hand, particles with $\beta$ $<$ 0.01 are also not considered because those grains move very slowly. Thus, simulating a significant mass flux of these particles would require to integrate the evolution of a huge number of super-particles or increase too much their mass. So, these simulations become prohibitive. 

For these experiments, we modeled the Poynting-Robertson drag in the {\tt mfo\_user} subroutine from the {\it Mercury-package} \citep{chambers1999}, so the dust grains could feel this drag. 

The dust grains were created in the code by randomly selecting parent massive planetesimals, one at a time, in the disk and turning part of its mass into a collection of dust super-particles. In other words, the mass of each selected planetesimal is decreased as $m'_p = m_p - k*m_{dust}$, where $k$ represents the number of dust super-particles created and $m_{dust}$ is their individual mass. The same planetesimal can be selected more than once, provided that $m_p > k*m_{dust}$. The parent planetesimal keeps its original position and velocity, and the dust grains are generated with the velocity of the parent body and just a small $x$-$y$ random displacement to avoid that they have subsequent identical evolutions. In fact, chaotic dynamics ensures that the super-particles will have independent evolutions for virtually any small relative displacement.  This procedure is adopted in order to avoid discontinuities in the time-distribution of the masses, positions and velocities in the disk,  which would occur if we created dust at random locations and/or without changing the planetesimal's mass. 

Following table \ref{table2}, the planetesimal disk was considered far enough from Neptune ($\delta >>\delta_{stab}$), with density and mass such that, without the generation of dust, the planets would not migrate and would be stable. In this way we can appreciate the effects of the dust on the dynamics of the planets.

We carried the experiments for three different values of $\beta$ (0.1, 0.05, and 0.01) also considering different dust fluxes ($\phi_{dust} =$ 1, 10, and 20 $M_{\oplus}$ per 400 My, hereafter $\phi_{dust}^{1}$, $\phi_{dust}^{10}$, and $\phi_{dust}^{20}$, respectively). A collection of $k$ dust grains was created every 10,000 years of integration, from a different planetesimal.  Thus, 2.5$\times 10^{-5}$, 2.5$\times 10^{-4}$, and 5.0$\times 10^{-4}$ $M_{\oplus}$ of dust were created every 10,000 years, respectively for the three values of $\phi_{dust}$ we investigate.
The mass of a single dust grain, then, is defined by the value of $k$. 
After several preliminary short time simulations ($t$ $<$ 50 My) considering $k = 1 ... 40$ (which means $100...4,000$ particles per My), we saw virtually the same evolution of Neptune.
This is expected because even the largest ``dust grain'' (5.0$\times 10^{-4}$ $M_{\oplus}$ for $k=1$ and $\phi_{dust}^{20}$) is still very small compared to Neptune's mass ($\sim$15 $M_{\oplus}$). 
What is important for the planet dynamics is not the size of each dust grain, but instead, its $\beta$ and the considered mass flux. So, for the main simulations we adopted $k=1$.

\begin{figure}
	\includegraphics[width=\columnwidth]{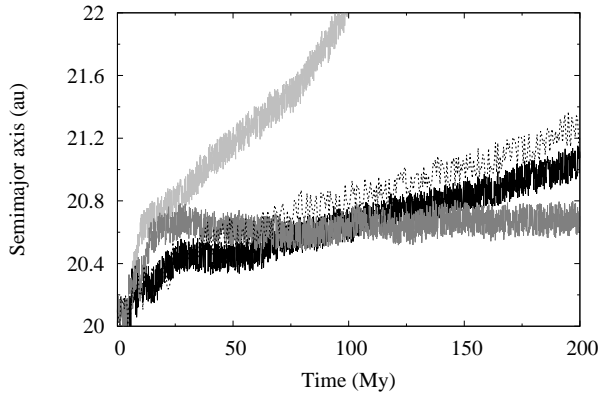}
    \caption{
Effects of $\beta$ over the evolution of Neptune's semimajor axis as a function of the time, for a fixed dust flux ($\phi_{dust}^{20}$) of 20 $M_{\oplus}$ per 400 My. 
Dark gray: $\beta=$ 0.1. 
Solid black: $\beta=$ 0.05.
Light gray: $\beta=$ 0.01.
Dashed black: $\beta$ uniformly distributed and randomly selected in the range between 0.1 and 0.01.
}
    \label{fig2}
\end{figure}

What was observed (figure \ref{fig2}) is: a first phase of a few My during which the planet does not migrate; a second phase of fast migration, independent of beta and mostly related to the distant encounters between the fifth and fourth planets after they have gone out of resonance, that brings the planet a few tenths of au away from its initial position in $\sim$20 My; then a third phase of slow, constant migration.  
This last phase is the one driven by the interaction with the dust flow. We observe that, for $\beta=$ 0.1, the planet basically does not migrate (dark gray in figure \ref{fig2}). For $\beta=$ 0.05 (black) and 0.01 (light gray) we clearly see outward migration, with a rate linearly anti-correlated with $\beta$.

Next, we ran several additional cases for dust fluxes $\phi_{dust}^{1}$, $\phi_{dust}^{10}$, and $\phi_{dust}^{20}$, and assigning a random value of $\beta$ between 0.1 and 0.01 with an uniform distribution for the dust grains created.  We think that this approach is valid and somewhat more realistic to represent a size-distribution of dust grains. 

These last runs resulted in an evolution very similar to those using $\beta =$ 0.05 (figure \ref{fig2} dashed black). This is because the existence of dust grains with $\beta$ close to 0.1 (ineffective in driving migration) is compensated by the existence of dust grains with $\beta \sim$ 0.01 (more effective).

\begin{figure}
	\includegraphics[width=\columnwidth]{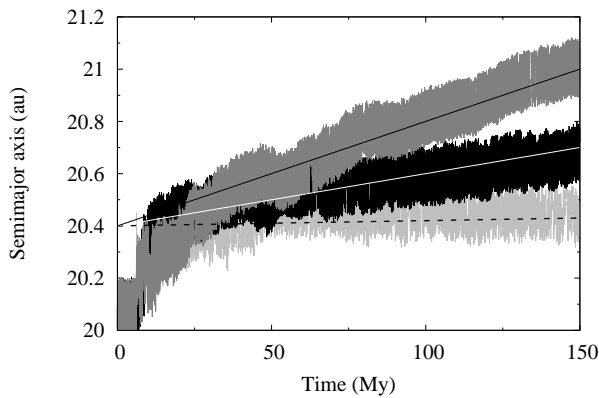}
    \caption{
Evolution of Neptune's semimajor axis under the effects of the dust flux ($\phi_{dust}$), as a function of the time. 
From bottom to top: {\it light gray} represents the evolution for  $\phi_{dust}^{1} =$ 1 $M_{\oplus}$ per 400 My, {\it black} for $\phi_{dust}^{10} =$ 10 $M_{\oplus}$ per 400 My, and {\it dark gray} for $\phi_{dust}^{20} =$ 20 $M_{\oplus}$ per 400 My. 
The straight lines, from bottom to top, represent one possible linear fit to these evolutions ({\it dashed black} for  $\phi_{dust}^{1}$, {\it white} for $\phi_{dust}^{10}$, and {\it solid black} for $\phi_{dust}^{20}$).
}
    \label{fig3}
\end{figure}

Adopting this prescription for $\beta$, figure \ref{fig3} shows the evolution of Neptune's semimajor axis under the effects of the dust flux ($\phi_{dust}$), as a function of the time. 
As in figure \ref{fig2}, one can immediately notice three changes in the migration regime: during the first $\sim$10 My Neptune does not migrate; then it migrates very rapidly up to $\sim$25 My; finally it migrates more slowly and linearly with time, with a rate proportional to the dust mass-flux as expected.  
 
In fact, in this third stage, Neptune's dust-driven evolution enters in a steady state regime. Each dust particle is first captured in MMR with Neptune, pushing the planet inwards, then is released and, through close encounters, moves the planet outwards.
The net effect of the dust is a slow outward migration of the planet, much slower than that obtained from an equivalent mass of planetesimals encountering the planet until their ultimate dynamical removal. 
This is because, at any given time, while some dust particles are having close encounters with Neptune pushing it outwards, some other particles are still in MMR with Neptune pushing it inwards (recall that, planetesimals don't push the planet inwards, even if they are captured in MMR because they don't feel the Poynting-Robertson drag). Moreover the lifetime of dust in the planet-encountering regime is shorter than for planetesimals (dust can drift across the planet's orbit without close encounters because of its radial migration).
 
In this phase the migration is linear in time and the migration rate depends linearly on the dust mass flux. We fit the averaged migration of Neptune with the formula 
$a_{_{N}}(\delta t) = a_{_{N,0}} + \gamma_{_{N}} \delta t$, with $a_{_{N,0}} \sim$20.4 au, and $\gamma_{_{N}} \sim$ 0.0002, 0.002, and 0.004 au/My for $\phi_{dust}^{ 1}$ (dashed black line), $\phi_{dust}^{10}$ (white line), and $\phi_{dust}^{20}$ (solid black line), respectively.
From the linear reletionship of $\gamma_{_{N}}$ on $\phi_{dust}$ one can easily estimate the migration of Neptune for any dust mass-flux. 

\subsection{Delayed evolution}\label{dde}

In this section we combine the results of sections \ref{dr} and \ref{de} to obtain a late planet instability starting from the configuration (3:2, 3:2, 2:1, 3:2), while fulfilling all constraints on the planetary evolutions (particularly criterion E presented in Sect.~\ref{early}).

The idea is the following. Under the effect of the dust flux, Neptune leaves its MMR with the adjacent planet in about 10 My and its orbital semimajor axis rapidly moves outwards by $\sim$0.4 au. If at this point the planet is still more than $\delta_{mig}$ away from the inner edge of the disk, the evolution of the planet will be dominated by dust-driven migration (with one caveat discussed below). If the dust flux has the appropriate magnitude, Neptune can reach the distance $\delta_{mig}$ from the disk after hundreds of My. When this point is reached, rapid planetesimal-driven migration will bring the planet to its final position in again $\sim$10 My and during this rapid migration the planetary instability will occur, as in sect.~\ref{early}.

The caveat that we anticipated above is that when the distance of the planet from the disk is larger than $\delta_{mig}$ but less than $\delta_{stab}$ some planetesimals start to interact closely with the planet. Thus, the migration of the planet is no longer driven only by the dust, but by a combination of dust and planetesimals effects. In absence of the dust, the planetesimal-driven migration would rapidly damp out, because the distance of the planet from the disk is larger than $\delta_{mig}$ and because of the very definition of this parameter (the largest distance from which planetesimal-driven migration is not in a damped mode). But, because the dust imposes a linear migration to the planet, planetesimals keep escaping from the disk through the moving MMRs with Neptune and add their contribution to the migration rate.

To evaluate this contribution, we proceeded with a series of simulations. We considered a disk with inner edge at 25 au, $\Sigma_0 =$ 0.03 $M_{\oplus}/au^2$, and $m_{disk} =$ 25.8 $M_{\oplus}$, modeled with 1,000, 5,000, and 10,000 planetesimals. We imposed a linear outward migration to Neptune  with the rates $\gamma_{_{N}}$ found in the previous section and we measured the actual  migration rate of the planet, resulting from the combination of the migration rate we impose and of close encounters with planetesimals escaping from the disk. 

With these series of 9 simulations (three resolutions and 3 values of $\gamma_{_{N}}$) we found that the actual migration rate $\gamma'_{_{N}}$ is about 2.8 times the imposed rate $\gamma_{_{N}}$, whatever the value of $\gamma_{_{N}}$ and the disk's resolution. Obviously the value 2.8 is likely linear on the surface density of the disk, but should also depend on the actual distance of the planet from the disk (i.e. which MMRs sweep the disk). But because the disk parameters used in these simulations are close to those of the disk that we will use below, we will adopt the factor 2.8 to set up our ultimate simulation.  

Our ultimate simulation will be a proof of concept that it is possible to obtain an instability delayed by about 300--400 My from the (3:2, 3:2, 2:1, 3:2) initial planetary configuration. The simulation accounts for the combined effects of dust and planetesimal scattering, with no imposed migration. We consider this simulation a ``proof of concept'' because a constant dust mass-flux is applied (a simplistic assumption; see next section) and the disk is quite under-resolved (1,000 planetesimals). But, given that a non zero value of $\delta_{stab}$ exists for whatever disk resolution, it is clear that with enough computational resources a similar evolution could be obtained for a more realistic disk resolution, rescaling the dust mass-flux in an appropriate way (see Table \ref{table3}). 
Also, according to \citet{izidoro2015}, if Uranus and Neptune formed from a set of proto-cores migrating inwards and stuck in resonance with Jupiter and Saturn, some gap has to be expected.

Using the information obtained in the previous sections, one can compute the dust flux needed in order to achieve an instability at a desired (late) time. For instance, let us suppose a disk (arbitrarily) with $\Sigma_0 =$ 0.03 $M_{\oplus}/au^2$ and an inner edge 5~au beyond the initial position of the planet, i.e., beyond the value $\delta_{stab}$ for a high resolution disk. We know that, after about 10~My of evolution, Neptune will be at about 20.4--20.6 au, namely will be $\sim$4.5 au away from the disk. We also know from sect~\ref{dr} that, for  the most resolved disk $\delta_{mig} \le$ 2 au, and that after Neptune reaches this distance, it takes about 10 My to the instability point. Thus we need that the planet migrates 2.5~au (i.e. from 4.5 au to 2 au from the disk), in say, 380 My.  This implies $\gamma'_{_{N}}=2.5/380$~au/My and, from the simulations above and the results of sect.~\ref{de}, $\gamma_{_{N}}=0.9/380$~au/My (where $\gamma_{_{N}}=\gamma'_{_{N}}$/2.8) and $\phi_{dust}\sim$11.5 $M_{\oplus}$ over 400 My. The values of  $\phi_{dust}$ for other initial planet-disk separations, and considering the value of $\delta_{mig}$ for the large disk resolution limit, are reported in table \ref{table3}. Only separations $\ge 4$~au are considered because those $\le 3$~au would lead to an early migration. 
 
In our proof of concept simulation, however, we can only deal with a low resolution disk due to computational limitations.
We use a disk resolution of 1,000 planetesimals and we assume $\Sigma_0 = 0.06 ~M_\oplus/au^2$. According to table \ref{table2}, in this case $\delta_{stab}=7$~au. So, we place the inner disk's edge at this distance from Neptune.
Thus, repeating the calculations above for this distance between Neptune and the disk, we find $\phi_{dust} \sim$ 20 $M_{\oplus}/$400 My.
The values of $k$ and $\beta$ were chosen as: $k=$ 1, and $\beta$ uniformly distributed and randomly selected in the range between 0.1 and 0.01. Finally, to reduce the effects of the coarse resolution of the disk on Neptune's migration, every time a planetesimal becomes Neptune's crosser, it is replaced by 10 new planetesimals with one tenth of the mass of the parent planetesimal each. This results in a disk that, when in contact with Neptune could have a maximum of 10,000 planetesimals. 

\begin{table}
	\centering
	\caption{
	Dust flux ($\phi_{dust}$) as a function of the initial separation $\delta_{stab}$ between Neptune and the disk, for the initial configuration (3:2, 3:2, 2:1, 3:2), necessary to bring Neptune from 20.5 au to $\delta_{mig} \le$ 2 au in $\sim$380 My when considering disks with enough resolution. 
	}
	\label{table3}
	\begin{tabular}{cccccc} 
		\hline
$\delta_{stab}$ (au)  & \vline & 4 & 5 & 6 & 7 \\
		\hline
$\phi_{dust}$ ($M_{\oplus}/$400 My)  & \vline & 7.05 & 11.5 & 16.45 & 21.15  \\
		\hline
	\end{tabular}
\end{table}

\begin{figure*}
\includegraphics[width=15.cm]{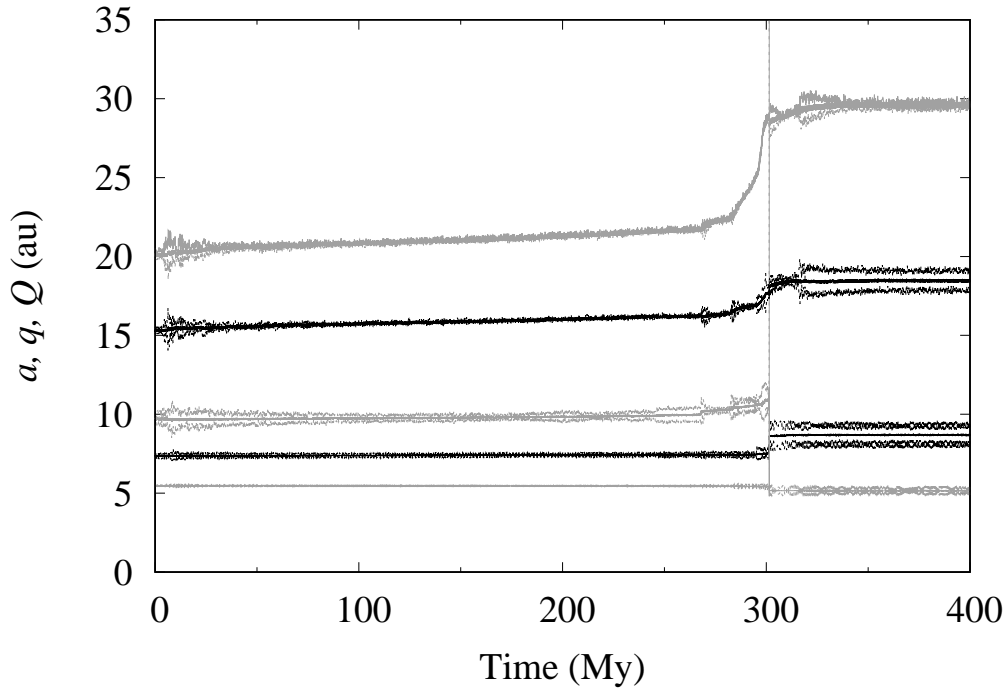}
    \caption{
Semimajor axis ($a$), perihelion distance ($q$) and aphelion distance ($Q$) for all planets as a function of the time. 
Planets are initially in the 3:2, 3:2, 2:1, 3:2 configuration with Jupiter at $\sim$5.4 au. 
$m_{disk} =$ 35 $M_{\oplus}$,  $\delta_{stab} =$ 7 au, with a resolution of 1,000 planetesimals initially. 
$\phi_{dust}=$ 20 $M_{\oplus}$ per 400 My, with $k =$ 1 and $\beta$ uniformly distributed and randomly selected in the range between 0.1 and 0.01.}
    \label{fig4}
\end{figure*}

\begin{figure*}
 \gridline{\fig{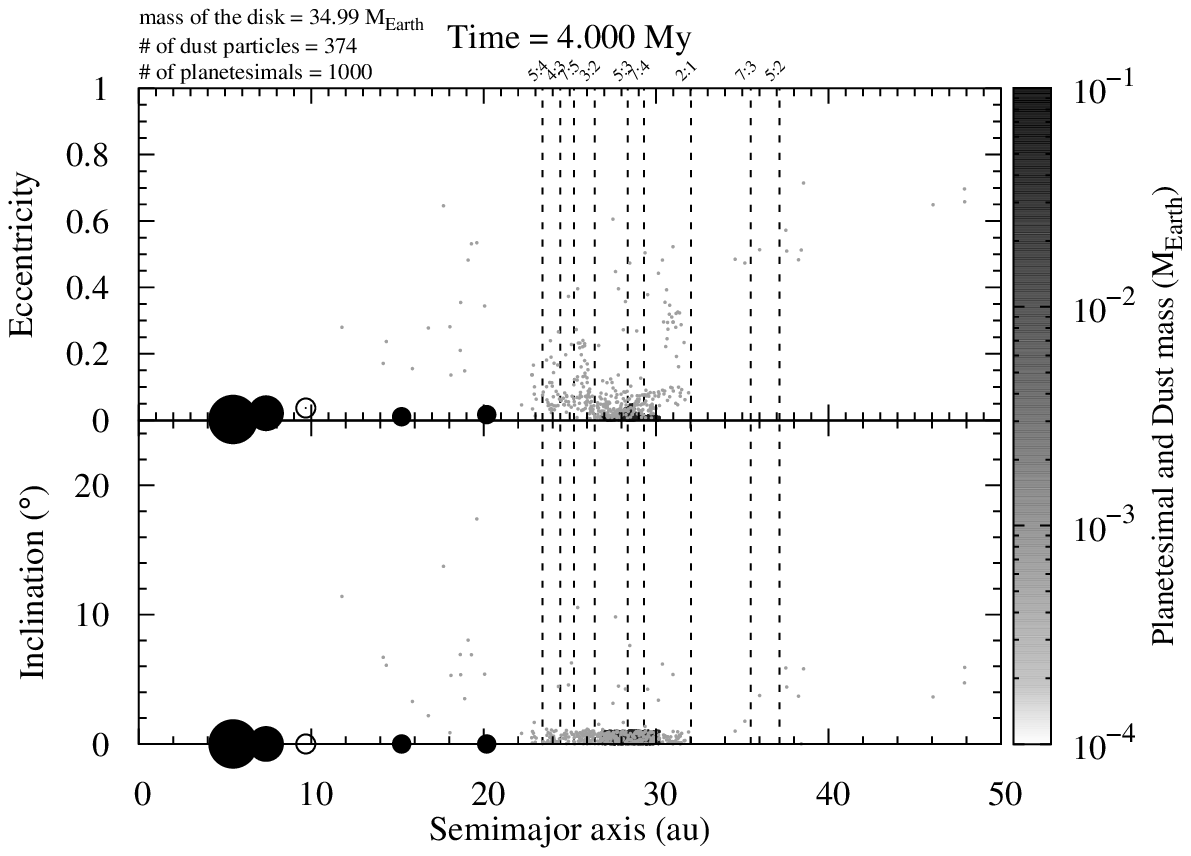}{.5\textwidth}{}
          \fig{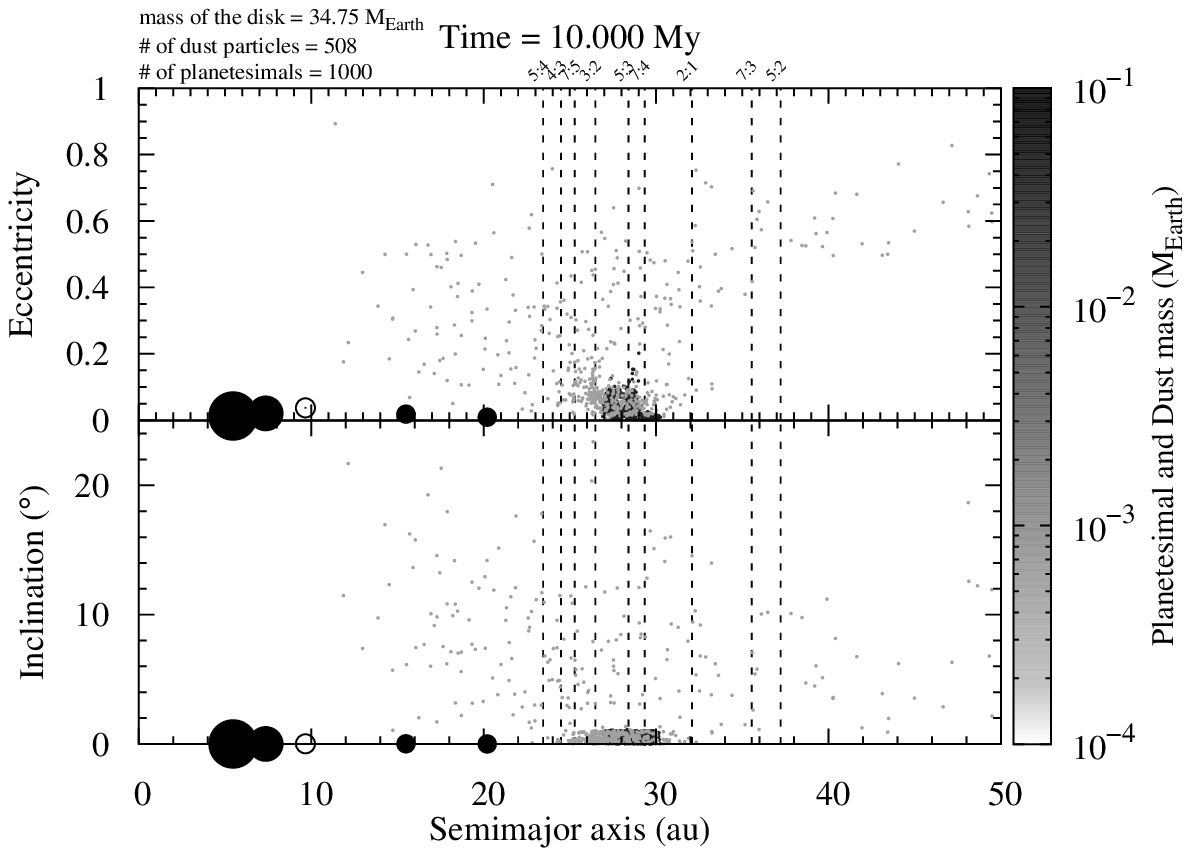}{.5\textwidth}{}}
\gridline{\fig{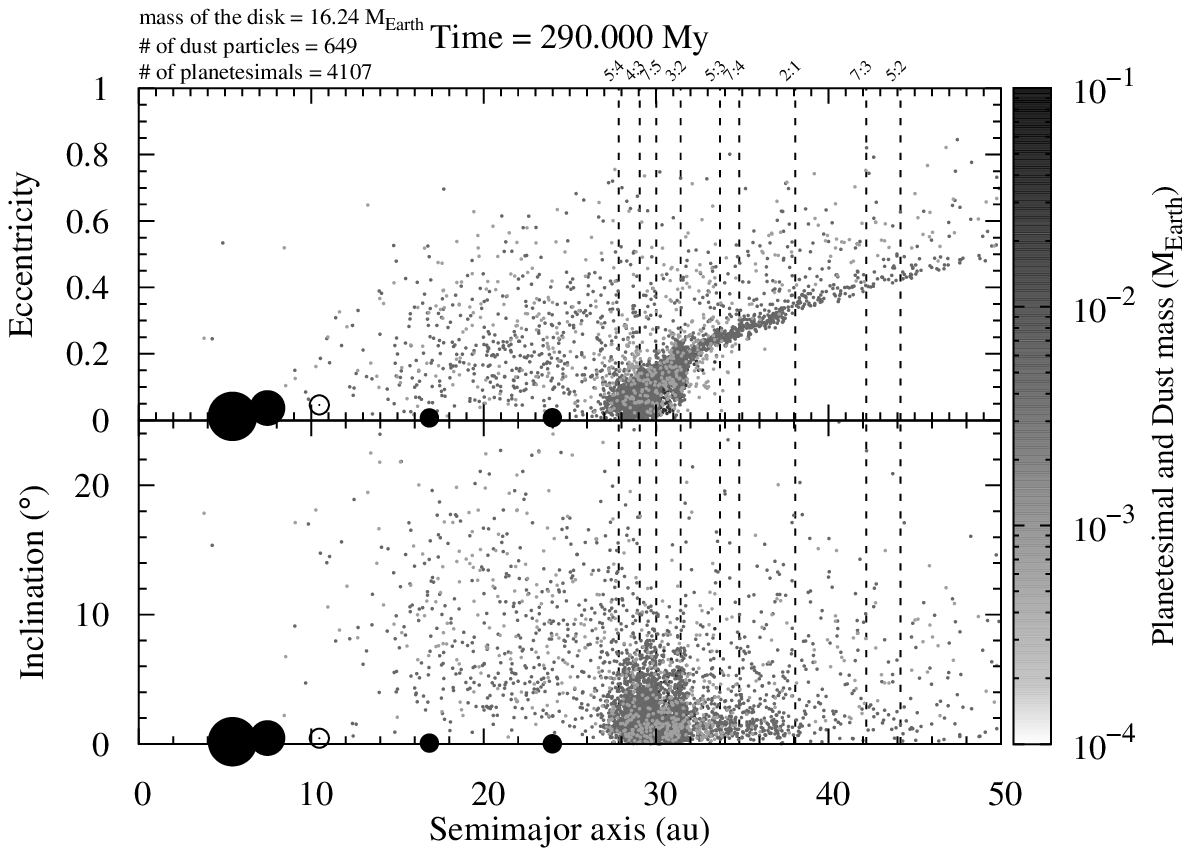}{.5\textwidth}{}
          \fig{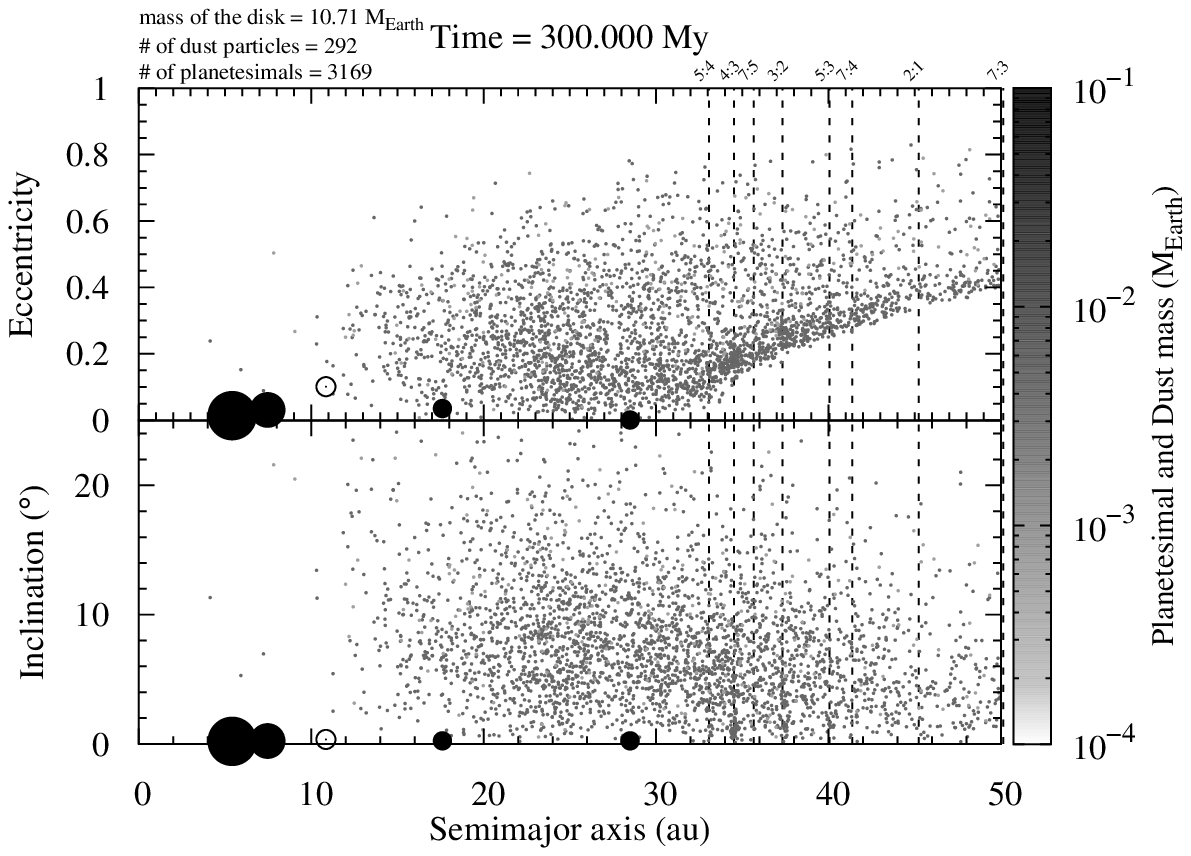}{.5\textwidth}{}}
\gridline{\fig{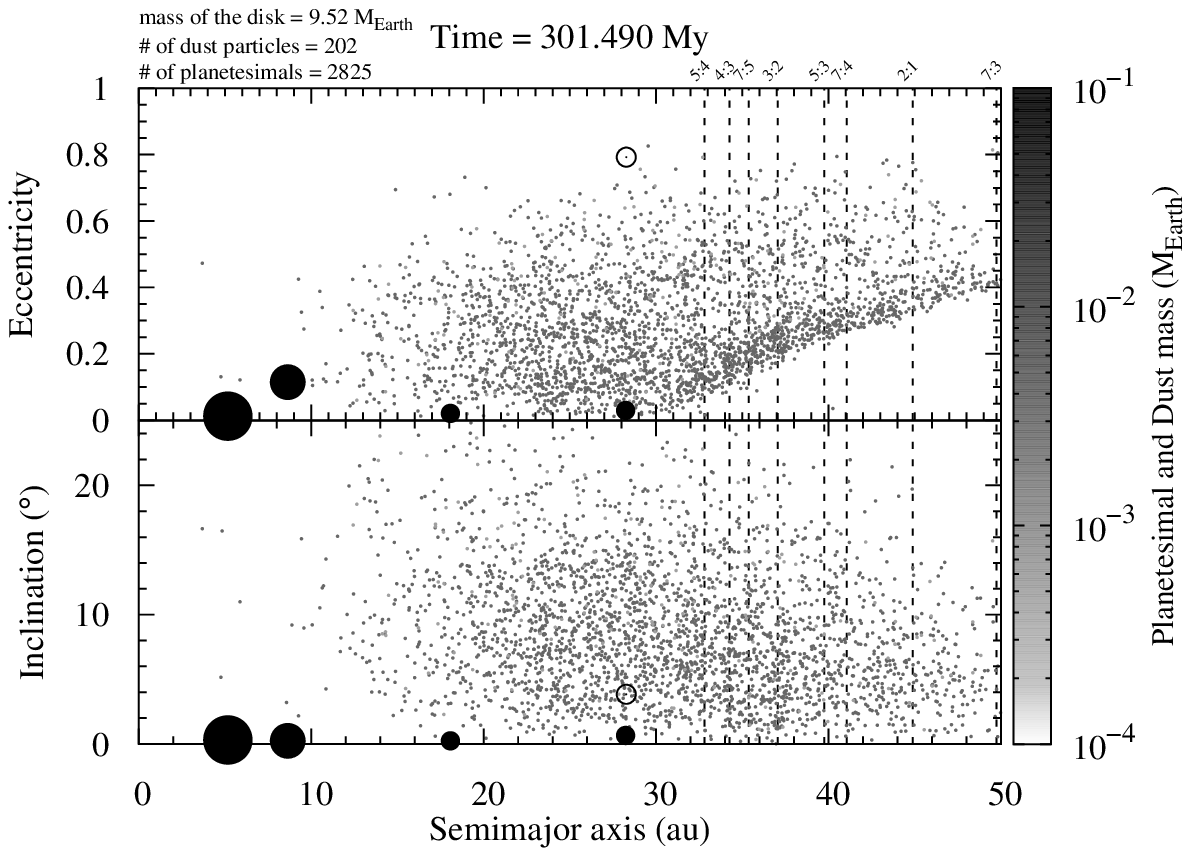}{.5\textwidth}{}
          \fig{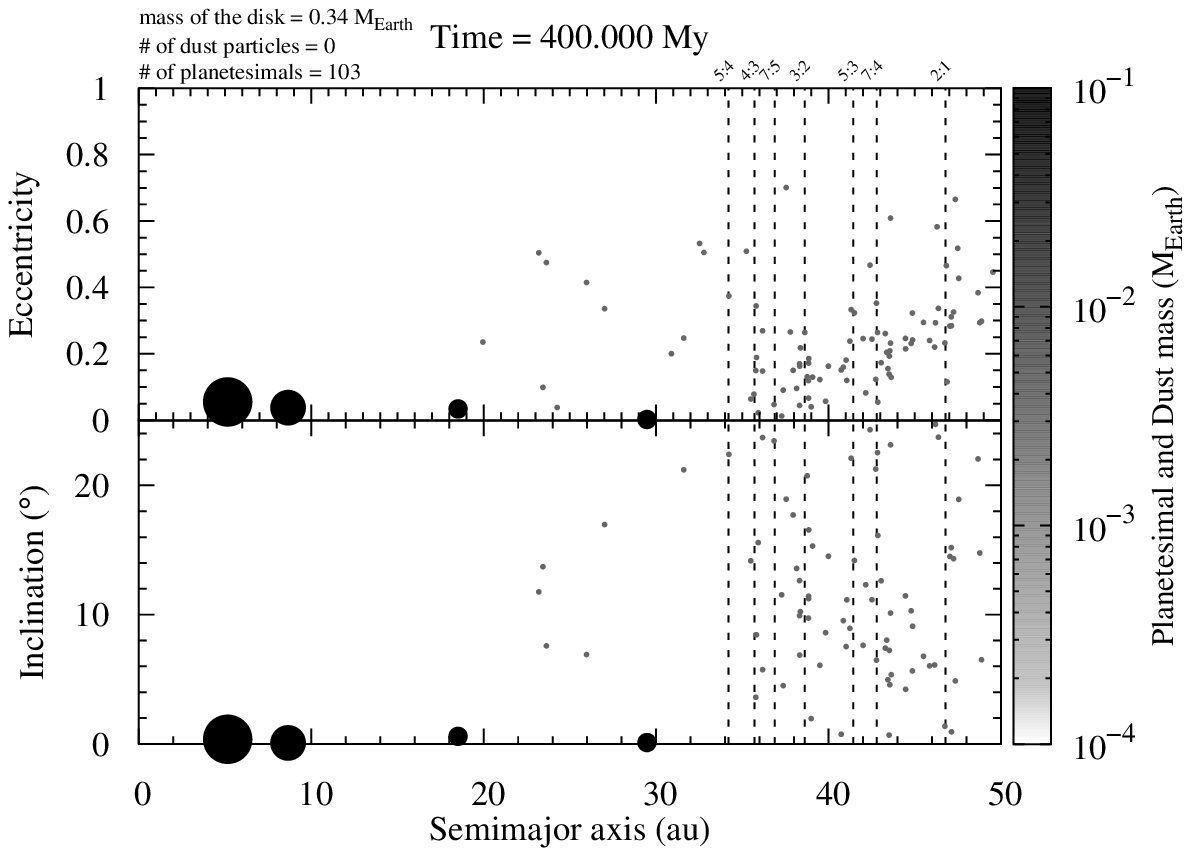}{.5\textwidth}{}}
   \caption{
The same as in figure \ref{fig4}, but here, the snapshots of the ($a$ $vs$ $e$) and ($a$ $vs$ $i$) of the whole system for different times of the evolution are shown. 
Jupiter, Saturn, Uranus, and Neptune are represented by filled black circles. 
The ejected fifht planet is marked as an open circle. 
The colored scale represents the individual mass of the planetesimals and of the dust grains present in the simulation at different times (the evolution of the planetesimal disk resolution is explained in the main text).
The vertical dashed lines show the location of the Neptune's MMRs into the planetesimal disk.
}
    \label{fig5}
\end{figure*}

The results of the ``proof of concept" simulation are shown in figures \ref{fig4} and \ref{fig5}. 
Figure \ref{fig4} shows the evolution of the semimajor axis ($a$), the perihelion distance ($q$) and the aphelion distance ($Q$) for all planets as a function of the time. 
Figure \ref{fig5} brings snapshots of the ($a$ $vs$ $e$) and ($a$ $vs$ $i$) of the whole system for different times of the evolution\footnote{An animation of the entire evolution can be found electronically at \href{url}{http://extranet.on.br/rodney/rogerio/delay-3.mp4}}.

The evolution shown in these figures has an instability after about 300~My and the evolution of the planets is in a considerably good agreement with that deduced in \citet{nesvorny2015a,nesvorny2015b} from Kuiper belt constraints. Particularly, the instability happens when Neptune is at $28$~au as required. 
Moreover, as we can see from these figures, the evolution of Neptune before the onset of the instability, follows closely the assumptions/predictions of tables \ref{table2} and \ref{table3}, and the dust mechanism of section \ref{de}. The instability did not happen precisely at the time we targeted, due to the approximations inherent in our linear combination of different processes, which in reality have a non-linear interplay. Nevertheless, the instability happens late enough to suggest that the association between a planetary instability and the trigger of the Late Heavy Bombardment is indeed possible.  

\subsection{Dust generation and lifetime}\label{dgl}

To assess whether the dust fluxes that are needed to migrate Neptune into the disk (see table \ref{table3}) are reasonable, we turned to the simulation presented in the supplementary material (S4) of \citet{levison2009}. That simulation considered a disk with initially 60 Earth masses, extended from 21 to 34 au. The disk was quite excited in eccentricity and inclination within 27 au and dynamically cold beyond it.

The planetesimals were assumed to have a size-frequency distribution (SFD) similar to that of the hot Kuiper belt population in the inner part of the disk (r $<$ 27 au) and similar to that of the cold Kuiper belt population beyond it. Both SFDs were scaled up in number so that the masses in the inner disk and outer disk were respectively 1/3 and 2/3 of the total. The collision probabilities and impact velocities within particles in the inner part of the disk, the outer part or in different parts were evaluated to be $3.8\times 10^{-21}$ km$^{-2}$y$^{-1}$ and 0.83 km/s, $2.5\times 10^{-21}$ km$^{-2}$y$^{-1}$ and 0.25 km/s, $4.2\times 10^{-22} $ km$^{-2}$y$^{-1}$ and 0.68 km/s, respectively.

The collisional evolution simulations were conducted with the code Boulder, developed in \citet{morbidelli2009}. The specific impact energy for collisional disruption was assumed to be 1/3 of that of strong ice \citep{benz1999,leinhardt2009}. According to these simulations, the disk produced 32 Earth masses of dust in 400 My. 
A total mass of 32 Earth masses is big compared to the numbers in table \ref{table3}. On the other hand, not all this dust would survive all the way to Neptune-crossing orbit, because the collisional cascade in the dust population comminute grains until they reach the ejection limit $\beta=0.5$. This part of the collisional cascade is not modeled in the Boulder code.

Also, the total amount of dust produced could be reduced by truncating the initial SFD to objects larger than some limit size;
in a new simulation where we retained only bodies with $D>75$ km in the initial SFD the total dust production is reduced to 15 Earth masses.  Another possibility is to increase the specific energy of disruption for the bodies. In fact, \citet{leinhardt2012} revised the previous estimates of \citet{leinhardt2009} and concluded that the specific energy for disruption should be similar to that for strong ice in \citet{benz1999}. A simulation with an initial truncated SFD  
but with a specific disruption energy 3 times larger produced only 7 Earth masses of dust.  Even more simply, reducing the initial population in the disk would reduce the amount of dust produced, given that the dust production scales as the square of the number of planetesimals in the disk. In conclusion, we think that the dust fluxes listed in table \ref{table3}, required to cause a significant migration of Neptune, are reasonable and in the good ballpark of what a massive planetesimal disk can produce.

A more relevant problem is the dust production timeline. In the Boulder experiments described above, the dust production decays as $1/t$. In the Boulder simulations, the collisional evolution of the dust is not tracked. But the dust collisional lifetime should scale as the square root of the dust production. Consequently, the total amount of dust of a given size (or $\beta$) should evolve with time as $1/\sqrt{t}$. Because the migration of the planet is linear in the amount of dust that it encounters, the evolution of its semimajor axis should be proportional to $\sqrt{t}$. This means that, of the total migration range spanned in -say- 400 My, 1/2 of it is covered in the first 100 My and 13\% in the last 100 My.  
In other words, the time to reach $\delta_{mig}$ is proportional to $(\Delta a_{_N}/\phi_{dust})^2$, instead of being proportional to $\Delta a_{_N}/\phi_{dust}$ as in the case of a linear evolution (where $\Delta a_{_N}$ is $\sim \delta_{stab}-\delta_{mig}-0.5$ au). 
Thus, small changes in the flux can make the curve $a_{_N} \propto \sqrt{t}$ to reach $\delta_{mig}$ distance early or very late.
Obtaining a late instability around 400 My requires a more delicate tuning of the dust production than that done in sect. \ref{dde}, where we considered that the dust production is constant and the collisional lifetime of dust infinite. 

\section{Conclusions}\label{conclusions}

It is very likely that, at the disappearance of the protoplanetary disk of gas, the giant planets of the Solar System were all in mean motion resonances with each other, because this is the typical outcome of gas-driven planet migration \citep{morbidelli2007}. \citet{nesvorny2012} investigated which resonant configurations are more likely to reproduce the current orbital configuration of the giant planets via a phase of dynamical instability, as that described in the Nice model. They considered four criteria of success and found a number of initial configurations that meet all of them in a non-negligible fraction of the simulations. All these configurations require the existence of a fifth giant planet, with a mass comparable to those of Uranus or Neptune, eventually ejected during the instability.

More recently, \citet{nesvorny2015a,nesvorny2015b} showed that explaining and reproducing the current orbital structure of the Kuiper belt requires that Neptune had a specific evolution, reaching a heliocentric distance of $\sim$28 au by planetesimal-driven migration before that the instability happened.

Thus, in section \ref{early} of this paper, we have revisited the work of \citet{nesvorny2012}, adding a new criterion of success from the results of \citet{nesvorny2015a,nesvorny2015b}. Of the configurations selected by \citet{nesvorny2012}, the one denoted 3:2, 3:2, 2:1, 3:2 (each integer ratio referring to the mean motion resonance between a pair of adjacent planets) stands out as the more likely to fulfill the new criterion of success. Other configurations (the 3:2, 3:2, 2:1, 2:1, the  3:2, 3:2, 3:2, 3:2, and the 3:2, 3:2, 4:3, 4:3) cannot be excluded, but their success rate relative to the new criterion is significantly lower.

An open issue with the Nice model is the timing of the giant planet instability. If one wants to explain the Late Heavy Bombardment of the Moon as a spike in the impact crater rate \citep{tera1974}, the instability has to have happened late \citep{gomes2005,levison2011,bottke2012}, unlike in the simulations of section \ref{early}, in which the instability occurred after a few 10s of My. It is not obvious that a 5-planet resonant configuration can remain stable for long time and become unstable then, particularly with a specific evolution of Neptune as that required to sculpt the Kuiper belt. Thus, in section \ref{late} we investigated whether a late instability from the 3:2, 3:2, 2:1, 3:2 is possible and under which conditions.
First, we looked for characteristics of the planetesimal disk that can maintain the planets in stable orbits for 400 My. We found that the evolution and stability of the planets are very sensitive to the resolution of the planetesimal disk. 
Thus, we determined the minimum distance that the inner edge of the planetesimal disk has be away from Neptune to keep the planets in stable orbit for 400 My as a function of the surface density of the disk and its numerical resolution. We named this distance $\delta_{stab}$ (table \ref{table2}).
In a second step, we found a characteristic Neptune-disk distance, which we named $\delta_{mig}$, from where Neptune starts a long-range planetesimal-driven migration. Thus, for the system to exhibit a late instability together with an evolution of Neptune like the one that is needed to sculpt the Kuiper belt, Neptune needs to be extracted from the 3:2, 3:2, 2:1, 3:2 resonance chain, and to migrate outwards over hundreds of Mys until it reaches the distance $\delta_{mig}$. 
To understand how this could be possible, we introduced the concept of dust-driven migration, i.e., migration due to the gravitational interaction of a planet with the dust produced in a distant planetesimal disk, spiraling inward by Poyning-Robertson drag. We have shown that a late instability can be achieved when considering a flux of dust ($\phi_{dust}$) ranging from 7 to 22 $M_{\oplus}$ per 400 My, depending on the value of $\delta_{stab}$ (table \ref{table3}).

Continuing, in section \ref{dgl} we evaluated how much dust could be produced from the disk-like configurations that we assumed and found good agreement with the dust flux described above. Indeed, we found that the total amount of dust produced in 400 My may vary from $\sim$7 $M_{\oplus}$ to a maximum of about 32 $M_{\oplus}$ depending on the Size Frequency Distribution (SFD) and catastrophic disruption energies assumed for the planetesimal disk and its constituents. Finally, we also found that the dust production rate expected for a self-grinding disk is not constant, but it decays as $1/t$. 
The abundance of dust, which is also limited by the dust collisional lifetime, should decay as $1/\sqrt{t}$.
Thus, because the migration of the planet is linear in the amount of dust that it encounters, the evolution of its semimajor axis should be proportional to $\sqrt{t}$, which implies that small changes in the flux of dust can result in Neptune reaching $\delta_{mig}$ distance early or very late. Therefore, obtaining
 a late instability around 400 My requires a more delicate tuning of the dust production.

Summarizing, based on the work of \citet{nesvorny2012}, and considering the evolution of Neptune required to explain the structure of the Kuiper belt \citep{nesvorny2015a,nesvorny2015b}, our results indicate that the preferred initial configuration of the giant planets at the time of the dissipation of the gas nebula is the 3:2, 3:2, 2:1, 3:2, and that, 
although an early instability reproduces more easily the evolution of Neptune required to explain the structure of the Kuiper belt, such evolution is also compatible with a late instability.

\section*{Acknowledgements}

We thank an anonymous reviewer for useful comments and suggestions on the submitted manuscript. R.D. acknowledges support provided by grants $\#$2015/18682-6 and $\#$2014/02013-5, S\~ao Paulo Research Foundation (FAPESP) and CAPES. A.M. acknowledges support by the French ANR, project number ANR-13--13-BS05-0003-01  projet MOJO (Modeling the Origin of JOvian planets). R.S.G. acknowledges his grant no. 307009/2014-9 from CNPq, Conselho Nacional de Desenvolvimento Cient\'ifico e Tecnol\'ogico, Brazil. The work of D.N. was supported by the NASA Emerging Worlds program.



\begin{thebibliography}{99}
\bibitem[Batygin et al.(2012)]{batygin2012} Batygin, K., Brown, M.~E., \& Betts, H.\ 2012, \apjl, 744, L3 
\bibitem[Benz \& Asphaug(1999)]{benz1999} Benz, W., \& Asphaug, E.\ 1999, \icarus, 142, 5 
\bibitem[Bottke et al.(2012)]{bottke2012} Bottke, W. F., Vokrouhlick\'y, D., Minton, D., et al. 2012, \nat, 485, 78
\bibitem[Brasil et al.(2016)]{brasil2016} Brasil, P.~I.~O., Roig, F., Nesvorn{\'y}, D., et al.\ 2016, \icarus, 266, 142 
\bibitem[Brasser et al.(2009)]{brasser2009} Brasser, R., Morbidelli, A., Gomes, R., Tsiganis, K., \& Levison, F. H.\ 2009 \aap, 134, 1790
\bibitem[Burns et al.(1979)]{burns1979} Burns, J.~A., Lamy, P.~L., \& Soter, S.\ 1979, \icarus, 40, 1
\bibitem[Chambers(1999)]{chambers1999} Chambers, J. E.\ 1999, \mnras, 304, 793
\bibitem[Deienno et al.(2011)]{deienno2011} Deienno, R., Yokoyama, T., Nogueira, E. C., Callegari, N. Jr., \& Santos, M. T.\ 2011, \aap, 536, A57
\bibitem[Deienno et al.(2014)]{deienno2014} Deienno, R., Nesvorn\'y, D., Vokrouhlick\'y, D., \& Yokoyama, T. 2014, \aj,148, 25
\bibitem[Deienno et al.(2016)]{deienno2016} Deienno, R., Gomes, R.~S., Walsh, K.~J., Morbidelli, A., \& Nesvorn{\'y}, D.\ 2016, \icarus, 272, 114 
\bibitem[Fernandez \& Ip(1984)]{fernandez1984} Fernandez, J.~A., \& Ip, W.-H.\ 1984, \icarus, 58, 109 
\bibitem[Fernandez \& Ip(1996)]{fernandez1996} Fernandez, J. A., \& Ip, W. H.,\ 1996, \planss, 44, 431
\bibitem[Gomes et al.(2004)]{gomes2004} Gomes, R.~S., Morbidelli, A., \& Levison, H.~F.\ 2004, \icarus, 170, 492 
\bibitem[Gomes et al.(2005)]{gomes2005} Gomes, R. S., Tsiganis, K., Morbidelli, A., \& Levison, H. F.\ 2005, \nat, 435, 466
\bibitem[Izidoro et al.(2015)]{izidoro2015} Izidoro, A., Morbidelli, A., Raymond, S.~N., Hersant, F., \& Pierens, A.\ 2015, \aap, 582, A99
\bibitem[Kaib \& Chambers(2016)]{kaib2016} Kaib, N.~A., \& Chambers, J.~E.\ 2016, \mnras, 455, 3561 
\bibitem[Leinhardt \& Stewart(2009)]{leinhardt2009} Leinhardt, Z.~M., \& Stewart, S.~T.\ 2009, \icarus, 199, 542 
\bibitem[Leinhardt \& Stewart(2012)]{leinhardt2012} Leinhardt, Z.~M., \& Stewart, S.~T.\ 2012, \apj, 745, 79 
\bibitem[Levison et al.(2007)]{levison2007} Levison, H.~F., Morbidelli, A., Gomes, R., \& Backman, D.\ 2007, Protostars and Planets V, 669 
\bibitem[Levison et al.(2008)]{levison2008} Levison, H. F., Morbidelli, A., Vanlaerhoven, C., Gomes, R., \& Tsiganis, K.\ 2008, \icarus, 196, 258
\bibitem[Levison et al.(2009)]{levison2009} Levison, H.~F., Bottke, W.~F., Gounelle, M., et al.\ 2009, \nat, 460, 364 
\bibitem[Levison et al.(2011)]{levison2011} Levison, H.~F., Morbidelli, A., Tsiganis, K., Nesvorn\'y, D., \& Gomes, R.\ 2011, \aj, 142, 152
\bibitem[Nesvorn\'y et al.(2007)]{nesvorny2007} Nesvorn\'y, D., Vokrouhlick\'y, D., \& Morbidelli, A.\ 2007, \aj, 133, 1962
\bibitem[Nesvorn\'y(2011)]{nesvorny2011} Nesvorn\'y, D.\ 2011, \apj, 742, 22
\bibitem[Nesvorn\'y \& Morbidelli(2012)]{nesvorny2012} Nesvorn\'y, N. \& Morbidelli, A.\ 2012, \aj, 144, 117
\bibitem[Nesvorn\'y et al.(2013)]{nesvorny2013} Nesvorn{\'y}, D., Vokrouhlick{\'y}, D., \& Morbidelli, A.\ 2013, \apj, 768, 45
\bibitem[Nesvorn{\'y} et al.(2014a)]{nesvorny2014} Nesvorn\'y, D., Vokrouhlick\'y, D., \& Deienno, R.\ 2014, \apj, 784, 22
\bibitem[Nesvorn{\'y} et al.(2014b)]{nesvorny2014b} Nesvorn{\'y}, D., Vokrouhlick{\'y}, D., Deienno, R., \& Walsh, K.~J.\ 2014, \aj, 148, 52 
\bibitem[Nesvorn{\'y}(2015a)]{nesvorny2015a} Nesvorn{\'y}, D.\ 2015, \aj, 150, 73 
\bibitem[Nesvorn{\'y}(2015b)]{nesvorny2015b} Nesvorn{\'y}, D.\ 2015, \aj, 150, 68 
\bibitem[Nesvorn{\'y} \& Vokrouhlick{\'y}(2016)]{nesvorny2016} Nesvorn{\'y}, D., \& Vokrouhlick{\'y}, D.\ 2016, \apj, 825, 94 
\bibitem[Malhotra(1993)]{malhotra1993} Malhotra, R.\ 1993, \nat, 365, 819 
\bibitem[Malhotra(1995)]{malhotra1995} Malhotra, R.\ 1995, \aj, 110, 420
\bibitem[Morbidelli et al.(2005)]{morbidelli2005} Morbidelli, A., Levison, H. F., Tsiganis, K., \& Gomes, R.\ 2005, \nat, 435, 462
\bibitem[Morbidelli et al.(2007)]{morbidelli2007} Morbidelli, A., Tsiganis, K., Crida, A., Levison, F. H., \& Gomes, R.\ 2007, \apj, 134, 1790
\bibitem[Morbidelli et al.(2009)]{morbidelli2009} Morbidelli, A., Bottke, W.~F., Nesvorn{\'y}, D., \& Levison, H.~F.\ 2009, \icarus, 204, 558 
\bibitem[Morbidelli et al.(2010)]{morbidelli2010} Morbidelli, A., Brasser, R., Gomes, R., Levison, H. F., \& Tsiganis, K.\ 2010, \aj, 140, 1391
\bibitem[Moro-Martin(2013)]{moro2013} Moro-Martin, A.\ 2013, Planets, Stars and Stellar Systems.~Volume 3: Solar and Stellar Planetary Systems, 431 
\bibitem[Petit et al.(2011)]{petit2011} Petit, J.-M., Kavelaars, J. J., Gladman, B. J., et al. 2011, \aj, 142, 131.
\bibitem[Pierens \& Nelson(2008)]{pierens2008} Pierens, A., \& Nelson, R.~P.\ 2008, \aap, 482, 333 
\bibitem[Roig \& Nesvorn{\'y}(2015)]{roig2015} Roig, F., \& Nesvorn{\'y}, D.\ 2015, \aj, 150, 186 
\bibitem[Tera et al.(1974)]{tera1974} Tera, F., Papanastassiou, D.~A., \& Wasserburg, G.~J.\ 1974, Earth and Planetary Science Letters, 22, 1 
\bibitem[Thommes et al.(1999)]{thommes1999} Thommes, E.~W., Duncan, M.~J., \& Levison, H.~F.\ 1999, \nat, 402, 635 
\bibitem[Toliou et al.(2016)]{toliou2016} Toliou, A., Morbidelli, A., \& Tsiganis, K.\ 2016, \aap, 592, A72 
\bibitem[Tsiganis et al.(2005)]{tsiganis2005} Tsiganis, K., Gomes, R. S., Morbidelli, A., \& Levison, H. F.\ 2005, \nat, 435, 459
\bibitem[Walsh \& Morbidelli(2011)]{walshmorby2011} Walsh, K.~J., \& Morbidelli, A.\ 2011, \aap, 526, A126 
\bibitem[Walsh et al.(2011)]{walsh2011} Walsh, K. J., Morbidelli, A., Raymond, S. N., O'Brien, D. P. \& Mandell, A. M.\ 2011, \nat, 475, 206

\end{thebibliography}
\end{document}